%% file: main.tex
\PassOptionsToPackage{dvipsnames}{xcolor}
\documentclass[aps,prl,reprint,superscriptaddress,longbibliography]{revtex4-2} 

\usepackage[normalem]{ulem}
\usepackage{xcolor}   
\usepackage{amsmath,amsthm,latexsym,amssymb,amsfonts}
\usepackage[english]{babel}
\usepackage[utf8]{inputenc}			
\usepackage{graphicx, texdraw}  
\usepackage{color}
\usepackage{bm}
\usepackage{enumerate}
\usepackage[linktocpage,colorlinks=true,linkcolor=blue,citecolor=blue,breaklinks=true,urlcolor=blue]{hyperref}
\usepackage{csquotes} 
\usepackage{color}
\usepackage{amsmath}
\usepackage{amsfonts}
\usepackage{amssymb}

\input{commands}

\renewcommand\vec{\bm}

\AtBeginDocument{
    \newwrite\bibnotes
    \def\bibnotesext{Notes.bib}
    \immediate\openout\bibnotes=\jobname\bibnotesext
    \immediate\write\bibnotes{@CONTROL{%
    apsrev41Control,author="08",editor="1",pages="1",title="0",year="1"}}
     \if@filesw
     \immediate\write\@auxout{\string\citation{apsrev41Control}}%
    \fi
}

\begin{document}


\title{Tuning upstream swimming of micro-robots by shape and cargo size}


\author{Abdallah Daddi-Moussa-Ider}
\thanks{All authors contributed equally to this work and are joint first, last, and corresponding authors.}
\email{abdallah.daddi.moussa.ider@uni-duesseldorf.de}
\email{mklis@fuw.edu.pl}
\email{amath@stanford.edu}
\affiliation{Institut f\"{u}r Theoretische Physik II: Weiche Materie, Heinrich-Heine-Universit\"{a}t D\"{u}sseldorf, Universit\"{a}tsstra\ss e 1, D\"{u}sseldorf 40225, Germany}

\author{Maciej Lisicki}
\thanks{All authors contributed equally to this work and are joint first, last, and corresponding authors.}
\email{abdallah.daddi.moussa.ider@uni-duesseldorf.de}
\email{mklis@fuw.edu.pl}
\email{amath@stanford.edu}
\affiliation{Institute of Theoretical Physics, Faculty of Physics, University of Warsaw, Pasteura 5, 02-093 Warsaw, Poland}

\author{Arnold J.T.M. Mathijssen}
\thanks{All authors contributed equally to this work and are joint first, last, and corresponding authors.}
\email{abdallah.daddi.moussa.ider@uni-duesseldorf.de}
\email{mklis@fuw.edu.pl}
\email{amath@stanford.edu}
\affiliation{Department of Bioengineering, Stanford University, 443 Via Ortega, Stanford, CA 94305, USA}
\affiliation{Department of Physics \& Astronomy, University of Pennsylvania, 209 South 33rd Street, Philadelphia, PA 19104, USA}

\begin{abstract}
The navigation of micro-robots in complex flow environments is controlled by rheotaxis, the reorientation with respect to flow gradients. Here we demonstrate how payloads can be exploited to enhance the motion against flows. Using fully resolved hydrodynamic simulations, the mechanisms are described that allow micro-robots of different shapes to reorient upstream. We find that cargo pullers are the fastest at most flow strengths, but pushers feature a non-trivial optimum as a function of the counter flow strength. Moreover, the rheotactic performance can be maximised by tuning the micro-robot shape or cargo size. These results may be used to control micro-swimmer navigation, but they also apply to rheotaxis in microbial ecology and the prevention of bacterial contamination dynamics.
\end{abstract}
%
%
%

\date{25 August 2020}

\maketitle

\section{Introduction}

For unicellular microorganisms, motility is an essential feature of life \cite{purcell1977}. To overcome or benefit from the fluid drag forces, these microbes have devised numerous swimming strategies \cite{Lauga2009}. Besides rich collective dynamics \cite{koch2011collective, elgeti2015, bechinger2016active}, even for isolated swimmers hydrodynamics can gravely affect microbial life \cite{wheeler2019not, rusconi2015}, through shape anisotropy \cite{degraaf16}, surface trapping \cite{berke08}, circular motion \cite{Lauga2006}, boundary accumulation \cite{spagnolie12, mino2018} and shear-induced accumulation \cite{Kessler1985, Rusconi2014} and swimming reorientation \cite{Barry2015, Luzio2005, DeGraaf2016}. 
Some microorganisms have also evolved to respond to flows, such as \textit{N. scintillans} dinoflagellates who exhibit bioluminescence to reduce grazing by predators that generate flows \cite{maldonado2007} and \textit{S. ambiguum} ciliates who perform hydrodynamic communication \cite{mathijssen2019collective}. However, so far only circumstantial evidence exists concerning the behavioural response to flow \cite{chengala2013}. It is therefore important to elucidate the inherent hydrodynamic mechanisms at play in microbiology. 

In a bulk shear flow, one of the sources for the complex behaviour is the geometry of the cells. Classical Jeffery orbits \cite{Jeffery1922} of elongated particles also apply to swimmer dynamics \cite{Zoettl2012}, as seen in experiments with {\it E. coli} bacteria~\cite{junot2019bacterium}. The chirality of their flagella was also shown to induce cross-streamline migration \cite{marcos2009separation, Marcos2012, jing2020chirality}. Interestingly, a rheotactic response leading to upstream swimming in bulk flows can also arise from viscoelasticity \cite{Mathijssen2016}.

Conversely, surfaces are known to alter hydrodynamic interactions in their vicinity, thus affecting the shear response significantly even for rigid particles~\cite{bretherton62}. Surfaces may enhance rheotaxis by providing a strong environmental coupling in which swimmers react to an external shear flow by orienting upstream. In particular, shear has been argued to aid navigation in mammalian sperm cells~\cite{Miki2013, Kantsler2014, Tung2015}, and govern the contamination dynamics of bacteria in channel flows~\cite{Hill2007, Kaya2012, Figueroa-Morales2015}. 
The dominant mechanism behind this upstream reorientation, termed the `weathervane effect', relies on the anisotropic and distance-dependent drag forces of the swimmer close to the surface. Far from walls this effect vanishes, which was also confirmed numerically \cite{uspal2015rheotaxis, ishimoto2017guidance}.
Even though this mechanism of surface rheotaxis is fairly understood, studies that couple this knowledge with other effects like confined Jeffery orbits, hydrodynamic wall attraction and chirality still lead to discoveries like oscillatory rheotaxis \cite{mathijssen2019oscillatory} and long-tailed distributions of run-tumble dynamics that can cause `super-contamination' \cite{figueroa2020coli}.

\begin{figure*}
		\includegraphics[width = \linewidth]{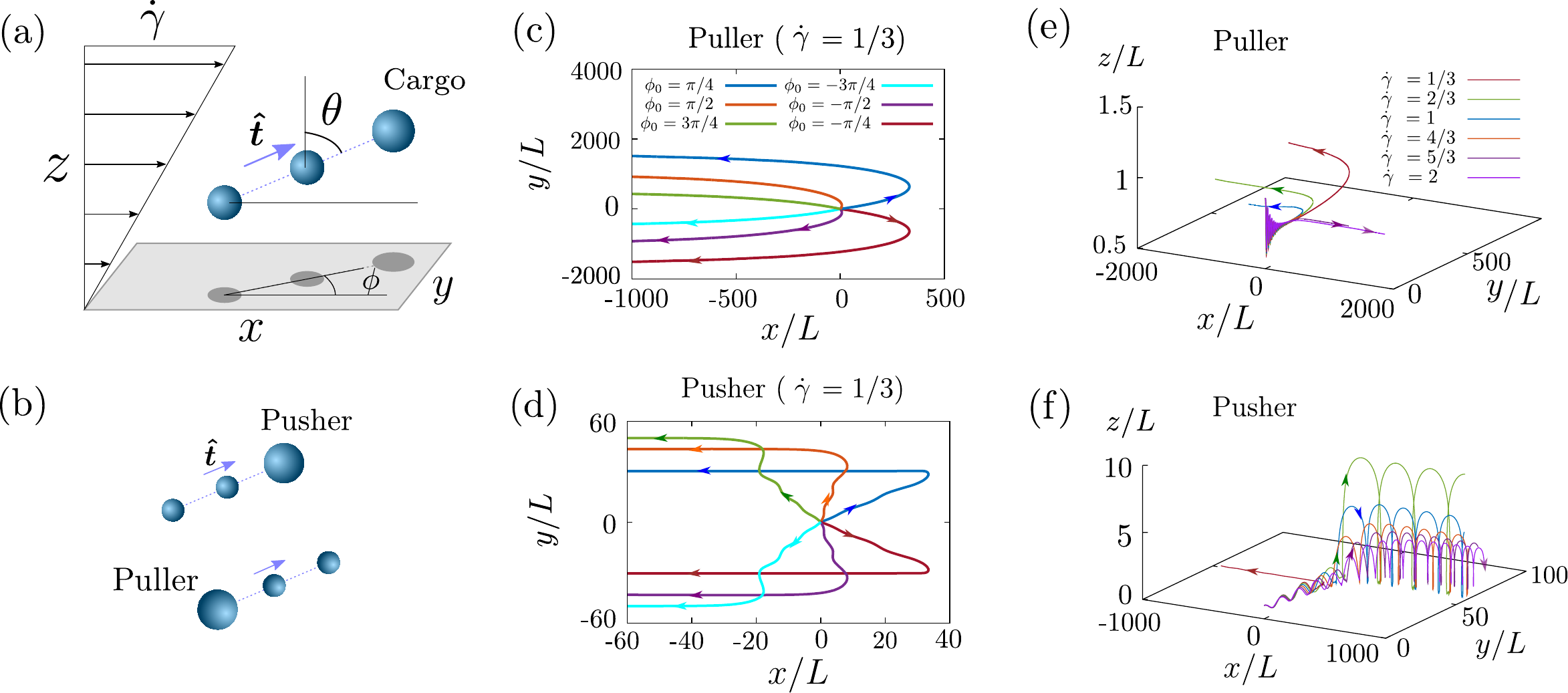}
	\caption{\label{Fig1}
    	\textit{\textbf{Surface rheotaxis in three dimensions.}}
    	\textbf{(a)}. 
    	Diagram of a three-sphere swimmer in shear flow near a surface. Shown is a pusher, with the large sphere at the front. 
    	\textbf{(b)}. 
    	Geometry of cargo pushers and pullers.
    	\textbf{(c,d)}. 
    	Swimming trajectories of $(c)$ pullers and $(d)$ pushers at various initial orientations $\phi_0$. The swimmers are initially released from $z_0=1$ and parallel to the surface, $\theta_0=\pi/2$.
    	\textbf{(e,f)}. 
    	3D trajectories at various shear rates. The swimmers are again released from $z_0=1$ with orientations $\phi_0=\theta_0=\pi/2$.
	}
\end{figure*}

Understanding the influence of flow on microorganism behaviour has opened the exploration of artificial rheotaxis, using synthetic nanoparticles and micro-robots. For these, upstream swimming in response to shear was also observed in a variety of contexts and for different propulsion mechanisms, including chemical and acoustic effects \cite{ren2017rheotaxis}, photocatalytic autophoretic systems of colloidal rollers \cite{Palacci2015} and rod-shaped Janus particles \cite{brosseau2019relating, baker2019fight}. A generic swimming mechanism for natural swimmers involving elastohydrodynamic coupling is also strongly related to the dynamics of the environment and flow conditions~\cite{Huang2019}.

In this contribution, we explore the transport of cargo by a model Najafi-Golestanian swimmer~\cite{najafi04, golestanian08} in an external shear flow close to a planar boundary, where one sphere is larger to hold the payload. Depending on the swimmer type (cargo pusher or puller), we observe different reorientation mechanisms that all lead to a positive rheotactic response. 
Hence, after reorienting upstream, the full three-dimensional dynamics reduce to a two-dimensional motion in the shear plane. 
This allows us to quantify the swimmer dynamics in a phase space spanned by the wall-separation distance and the head orientation.
By analysing the fixed points in these phase diagrams we identify the rheotactic states, and their stability for the different swimmer geometries.
Next, we map out the upstream migration speed as a function of imposed shear rate, and find that pushers and pullers perform optimally in completely different external flow conditions.
Finally, the rheotactic performance is tuned by regulating the cargo size at different flow strengths.

\section{Model}

We consider the dynamics of a neutrally buoyant micro-robot subject to an externally applied shear flow near a planar no-slip boundary in a Newtonian fluid.
As a model swimmer, we employ the linear three-sphere micro-swimmer originally proposed by Najafi and Golestanian~\cite{najafi04, golestanian08}, as schematically illustrated in Fig.~\ref{Fig1}(a).
More generally, our results apply to a broader class of micro-robots at low Reynolds numbers \cite{purcell1977}, so with a size smaller than  $\sim 1$mm, and with a swimming speed smaller than $\sim1$mm/s. 
Additionally, we consider the deterministic limit without translational and rotational diffusion. 
This is relevant when the (rotational) P\'eclet number is large, $\text{P\'e}_r > 1$, for micro-robots larger than $\sim 10\mu$m in size.

Throughout the paper, all quantities are non-dimensionalised by scaling lengths with the mean swimmer arm length, $L$, and velocities are scaled by the inverse of the free swimming speed in the absence of external flows and boundaries, $V_0$. The total mean length of the swimmer is thus $2L$.
The surface is located at $z=0$ in Cartesian coordinates, and the flow is given by $\vec{u} = \gamDot z \hat{\vec{x}}$ in terms of the shear rate $\gamDot$.
So, for clarity, the dimensional shear rate is $\gamDot^* = \gamDot V_0 / L$.
The swimmer is neutrally buoyant, and is composed of three spheres joined by thin arms, all aligned along the swimming direction, $\hat{\vec{t}}$.
The arm lengths oscillate with frequency $\omega$, respectively, at an angle $\pi/2$ out of phase.

We consider both cargo-pushing swimmers with a larger sphere at the front, and pullers with a cargo at the back [Fig.~\ref{Fig1}b]. 
The hydrodynamic signature, the far-field flow generated by such a three-sphere cargo pusher (puller) corresponds to an extensile (contractile) Stokes dipole \cite{dunkel10, Daddi2018state, nasouri2019efficiency}.
The radius of the two smaller spheres is $a = 0.1$ and the larger sphere has radius $a_+ = 0.12$ unless mentioned otherwise. Particularly later in the paper, where we will tune the upstream motility as a function of the cargo size.


\section{Simulation methods}

The swimming dynamics are found by solving for the hydrodynamic interactions between the spheres and the wall, including the external shear flow.
The swimmer is constituted of three spherical particles connected by dragless rods of negligible hydrodynamic effects to ensure their co-linearity.  
In order to achieve self-propulsion at low Reynolds numbers, micro-swimmers have to undergo a non-reciprocal sequence of shapes during their locomotion.
Accordingly, by periodically varying the mutual distance between the spheres in a non-reciprocal manner, a net swimming motion over one full cycle is achieved \cite{purcell1977}.

Owing to the linearity of the Stokes equations, the translational and rotational velocity of each sphere are related to the hydrodynamic forces and torques, respectively denoted as $\vect{F}$ and $\vect{L}$, via the generalized mobility tensor $\bmu$. 
Following Dhont \cite{dhontBook96}, these velocities in the laboratory (LAB) frame of reference are given by
    \begin{align} \nonumber
    \label{GeneralizedMobilityTensor}
    			\begin{pmatrix}
    				\vect{V}_\gamma \\
    				\bOmega_\gamma
    			\end{pmatrix}
    			&= 
    			\sum_{\lambda=1}^{3} 
    			\begin{pmatrix}
    				\bmu_{\gamma\lambda}^{tt} & \bmu_{\gamma\lambda}^{tr} \\
    				\bmu_{\gamma\lambda}^{rt} & \bmu_{\gamma\lambda}^{rr} 
    			\end{pmatrix}
    			\cdot
    			\begin{pmatrix}
    				\vect{F}_\lambda \\
    				\vect{L}_\lambda  
    			\end{pmatrix}
    			 \\ 
    			 &
    			+
    			\begin{pmatrix}
    				\vect{v}_\infty (\R_\gamma) \\
    				\bomega_\infty (\R_\gamma)
    			\end{pmatrix}
    			+
    			\begin{pmatrix}
    				\vect{C}_\gamma^t \\
    				\vect{C}_\gamma^r  
    			\end{pmatrix}
    			: \vect{E}_\infty \, , 
    \end{align} 
for $\lambda, \gamma  \in \{1,2,3\}$ being the sphere indices, where $\vect{v}_\infty (\R) = \vect{K}_\infty \cdot \vect{r}$ is the external shear flow, with $\vect{K}$ being the velocity gradient matrix, $\bomega_\infty (\R) = \frac{1}{2} \bNabla \times \vect{v}_\infty (\R)$ is the fluid vorticity, and $\vect{E}_\infty$ is the symmetric part of~$\vect{K}_\infty$.
Here, $\bmu^{\alpha\beta}$, with $\alpha \in \{t,r\}$ and $\beta \in \{t,r,d\}$, are the components of the hydrodynamic mobility tensor, where the superscripts $t$, $r$, and $d$ stand for the translational, rotational, and dipolar components, respectively.
Moreover, $\vect{C}^t$ and $\vect{C}^r$ denote the translational and rotational parts of the shear disturbance tensor, respectively, given by
    \begin{equation}
    	\vect{C}_\gamma^t = \sum_{\lambda=1}^{3} \bmu_{\gamma\lambda}^{td} \, , 
    	\qquad
    	\vect{C}_\gamma^r = \sum_{\lambda=1}^{3} \bmu_{\gamma\lambda}^{rd} \, .
    \end{equation} 
Our method uses an accurate representation of the solution of the mobility tensor of a sphere near the no-slip surface, following Cichocki \& Jones \cite{cichocki98}. This method includes near-field effects and lubrication, and is accurate at all separation distances between the particle and the boundary. 
The interaction between the spheres is modelled using the Rotne-Prager-Yamakawa approximation for different-sized particles, following Zuk \textit{et al.} \cite{zuk14}. 
Because the spheres are small compared to both the arm length and the oscillation amplitude, this approximation up to quadrupolar order is sufficient to capture the swimming dynamics, as is demonstrated in earlier works.

In order to undergo autonomous motion, the resultant of the forces and torques acting on the swimmer has to vanish.
Accordingly, we have
    \begin{equation}\label{forceUndTorqueFreiBedingung}
    		\sum_{\lambda=1}^{3} \vect{F}_\lambda = 0 \, , \quad\quad
    		\sum_{\lambda=1}^{3} (\R_\lambda-\R_0) \times \vect{F}_\lambda + \vect{L}_\lambda = 0 \, ,
    \end{equation}
where~$\R_0$ is a reference point for the torque moment calculation that, for convenience, we choose as the position of the central sphere, although any other choice would be just as good.

The instantaneous orientation of the swimmer relative to the wall is described by the unit vector $\hatt = \sin\theta \cos\phi \, \hatex + \sin\theta \sin\phi \, \hatey + \cos\theta \, \hatez$ pointing along the swimming direction.
Here, $\theta$ and $\phi$ represent the polar and azimuthal angles, respectively, in the spherical polar coordinates basis associated with the microswimmer.
In addition, we define the unit vectors $\hattheta = \cos\theta\cos\phi \, \hatex + \cos\theta\sin\phi \, \hatey-\sin\theta \, \hatez$ and $\hatphi = -\sin\phi \, \hatex + \cos\phi \, \hatey$, such that set of unit vectors $(\hatt, \hattheta, \hatphi)$ forms a direct orthonormal system.

\begin{figure*}[t]
    \includegraphics[width= \linewidth]{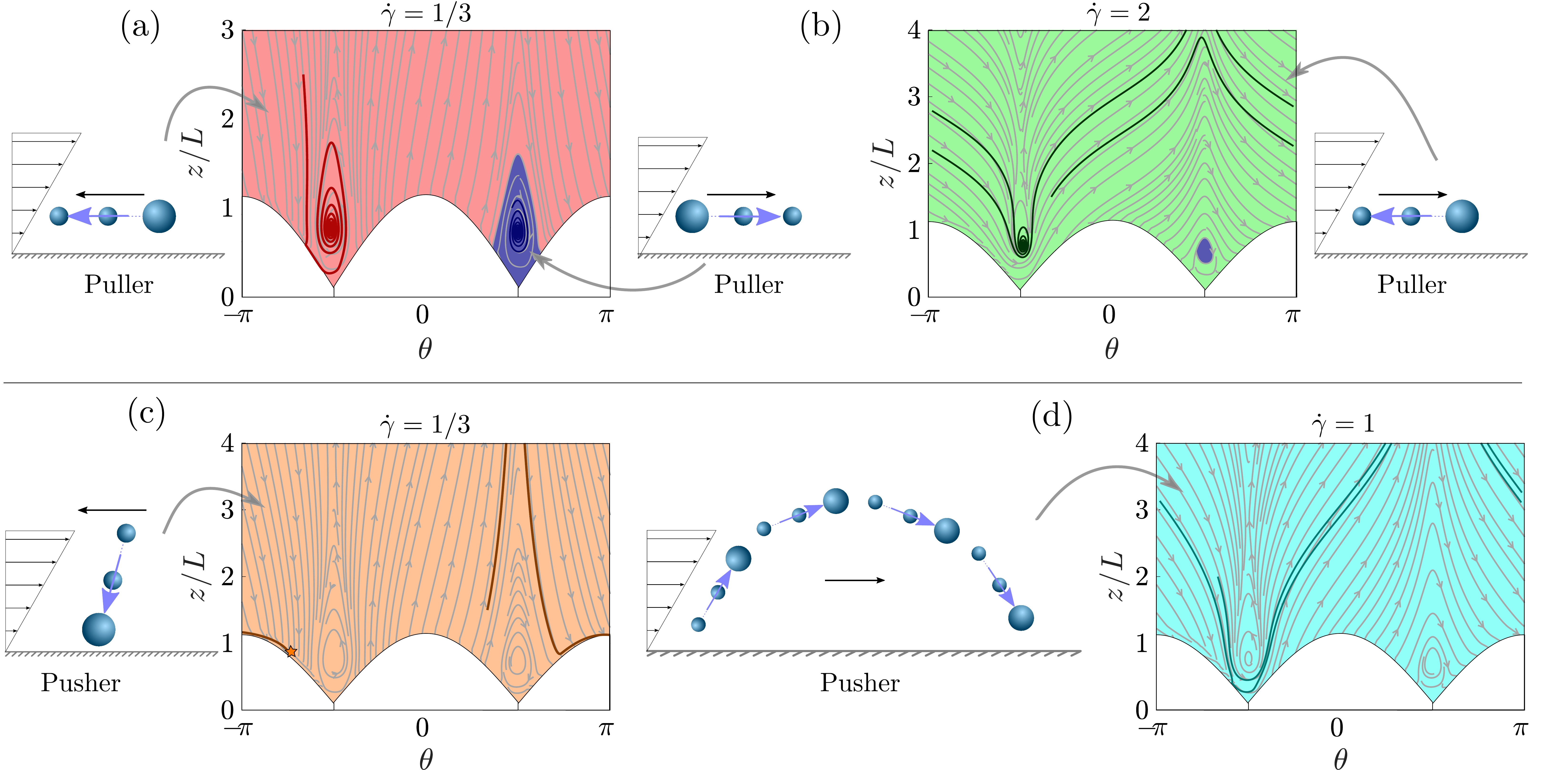}
    \caption{\label{Fig2}
        \textit{\textbf{Phase-space diagrams of upstream swimming,}} showing the  dynamics in $(\theta, z)$ space for various flow strengths. 
        Grey lines are streamlines in this phase space, coloured lines show example trajectories, and the background colours indicate the final state for each initial condition.
        Insets illustrate the corresponding final-state behaviours, as observed in real space, where blue arrows on the axis of the swimmer indicate its orientation $\hat{\vec{t}}$ and the black arrows above show the overall (lab frame) direction of motion, the sum of advection and self-propulsion. 
        \textbf{(a)}. Pullers at $\gamDot = 1/3$. 
        Red indicates that the swimmer ends up swimming upstream, parallel to the surface. 
        Blue is moving downstream, parallel to the surface.
        \textbf{(b)}. Pullers at $\gamDot = 2$. 
        Green indicates that the final state is moving downstream, but oriented upstream and parallel to the surface. 
        Blue as before.
        \textbf{(c)}. Pushers at $\gamDot = 1/3$. Brown shows that all swimmers move upstream, oriented almost perpendicular to the surface. The orange star indicates the final fixed point.
        \textbf{(d)}. Pushers at $\gamDot = 1$. Cyan shows that all swimmers are advected downstream, following an indefinite toppling motion detached from the surface. 
        The white arc-like regions are inaccessible due to the swimmer shape.
        }
\end{figure*}

In addition, we assume that the length of the rod joining the spheres varies periodically in time about a constant mean value. Specifically,
\begin{equation}\label{gUndhDefinitionen}
		\R_2 - \R_1 = g(t) \, \hatt \, , \qquad
		\R_1 - \R_3 = h(t) \, \hatt \, , 
\end{equation}
where $g$ and $h$ are harmonic functions that prescribe the instantaneous distances between the spheres. They are given by 
\begin{equation}
		g(t) = L + u_{0} \cos (\omega t) \, , \qquad
		h(t) = L + u_{0} \cos (\omega t + \delta) \, , 
\end{equation}
where $\omega$ denotes the oscillation frequency, $\delta \in (0,2\pi)$ is a phase shift that is required for symmetry breaking.
Moreover, $L$ is the mean length, and $u_{0}$ is the amplitude of periodic variations in the lengths of the rods, chosen small enough to ensure that $2|u_0| + a_1 + \max \{a_2, a_3\}\ll L$.

By combining the first row in Eq.~\eqref{GeneralizedMobilityTensor} providing the translational velocity of the $\gamma$th sphere with Eqs.~\eqref{gUndhDefinitionen} upon requiring that $\vect{V}_\gamma = \frac{\Intd \R_\gamma}{\Intd t}$, we obtain
    \begin{widetext}
    \begin{equation}\label{firstRowEq}
    			\sum_{\lambda=1}^{3}
    			\begin{pmatrix}
    				\vect{G}_\lambda^{tt} &  \vect{G}_\lambda^{tr} \\
    				\vect{H}_\lambda^{tt} &  \vect{H}_\lambda^{tr} 
    			\end{pmatrix}
    			\cdot
    			\begin{pmatrix}
    				\vect{F}_\lambda \\
    				\vect{L}_\lambda  
    			\end{pmatrix}
    			+ \begin{pmatrix}
    				\vect{C}_2^t - \vect{C}_1^t \\  
    				\vect{C}_1^t - \vect{C}_3^t \\  
    			\end{pmatrix}
    			:
    			\vect{E}_\infty 
    			= 
    			\begin{pmatrix}
    				\dot{g} \\
    				\dot{h}
    			\end{pmatrix}
    			\hatt 
    			+
    			\begin{pmatrix}
    			g \\
    			h
    			\end{pmatrix}
    			\left( \dot{\theta} \, \hattheta + \dot{\phi} \sin\theta \, \hatphi - \gamDot \cos\theta \, \hatex  \right) \, , 
    \end{equation}
    \end{widetext}
where we have defined, for convenience, the second-rank tensors
\begin{equation}
		\vect{G}_\lambda^{\alpha\beta} := \mu_{2\lambda}^{\alpha\beta}-\mu_{1\lambda}^{\alpha\beta} \, , \qquad
		\vect{H}_\lambda^{\alpha\beta} := \mu_{1\lambda}^{\alpha\beta}-\mu_{3\lambda}^{\alpha\beta} \, , 
\end{equation}
with $\alpha, \beta \in \{t,r\}$.
The reference frame associated with the swimmer can be obtained by performing two successive rotations following the standard Euler transformations~\cite{fossen11}, where~$\phi$ and $\theta$ correspond to the precession and nutation angles, respectively.
Accordingly, the angular velocities of the three spheres relative to the LAB frame are equal and are given by
\begin{equation}\label{angularVelocityEquation}
	\bOmega_\gamma = -\dot{\phi} \sin\theta \, \hattheta + \dot{\theta} \, \hatphi 
	+ \left( \dot{\phi} \cos\theta + \dot{\varphi} \right) \, \hatt  \, , 
\end{equation}
for $\gamma \in \{1,2,3\}$, where $\dot{\varphi}$ is an unknown (proper) rotation rate around the swimming axis.
It follows from the second row of Eq.~\eqref{GeneralizedMobilityTensor} expressing the angular velocities that
\begin{equation}\label{secondRowEq}
		\sum_{\lambda=1}^{3} 
		\begin{pmatrix}
			\vect{G}_\lambda^{rt} &  \vect{G}_\lambda^{rr} \\
			\vect{H}_\lambda^{rt} &  \vect{H}_\lambda^{rr} 
		\end{pmatrix}
		\cdot
		\begin{pmatrix}
			\vect{F}_\lambda \\
			\vect{L}_\lambda  
		\end{pmatrix}
		+ \begin{pmatrix}
			\vect{C}_2^r - \vect{C}_1^r \\  
			\vect{C}_1^r - \vect{C}_3^r \\  
		\end{pmatrix}
		:
		\vect{E}_\infty 
		= 
		0 \,  .
\end{equation}

By projecting Eqs.~\eqref{firstRowEq} onto the basis of spherical coordinates, and eliminating the unknown rotation rates~$\dot{\phi}$ and~$\dot{\theta}$, four scalar equations are obtained.
The projection of Eqs.~\eqref{forceUndTorqueFreiBedingung} and \eqref{secondRowEq} yields twelve additional equations. 
In addition, Eq.~\eqref{angularVelocityEquation} provides a closure of the linear system of equations by requiring that
\begin{subequations}
	\begin{align}
		\bOmega_1 \cdot \hattheta &= -\dot{\phi} \, \sin\theta \, , \\
		\bOmega_1 \cdot \hatphi   &= \theta \, .
	\end{align}
\end{subequations}
	
The determination of the eighteen unknown components of the internal forces and torques is thus achieved by solving the resulting linear system of equations using the standard substitution procedure. 
In order to obtain the swimming trajectories of the swimmer, we choose to track the instantaneous position of the central sphere along with the orientation of the swimming axis.
The positions of the front and aft spheres follow forthwith.
As mentioned in the main body of the paper, we scale all the lengths by the mean arm length~$L$, and the times by the inverse  of the oscillation frequency~$\omega$.

For the numerical computation of the swimming trajectories, we solve numerically the resulting dynamical system of equations using a standard Runge-Kutta scheme with adaptive time stepping~\cite{press92}.
In addition, we use tabulated results for the hydrodynamic mobility functions obtained using the exact multipole method of Stokes flows~\cite{cichocki00}.
Following the approach employed in Ref.~\cite{daddi18}, we include an additional soft repulsive force (excluded volume interactions) to avoid direct contact with the wall.

\section{Upstream swimming dynamics}

The three-dimensional dynamics of these pullers and pushers are first described for different initial orientations~$\hat{\vec{t}}_0$ parallel to the surface [Fig.~\ref{Fig1}c,d].
Indeed, we observe that all swimmers will eventually align with the shear plane, such that the component $\hat{\vec{t}} \cdot \hat{\vec{y}} \to 0$, for both swimmer types 
[also see Videos 1, 2].
This alignment also occurs for different shear rates [Fig.~\ref{Fig1}e,f].
As expected, stronger flows will reorient the swimmers more quickly.
Of course, at very strong shear the swimming speed no longer exceeds the local flow strength, leading to downstream advection 
[Videos 3, 4], 
but the swimmers can still be oriented upstream.


As a result of this alignment with the shear plane, the 3D trajectories reduce to two dimensions over time.
This is true in all tested cases, regardless of initial conditions, shear rate or swimmer type, as long as the swimmers come close enough to interact hydrodynamically with the surface. 
Then, the orientation of the swimmer in the shear plane is given by the pitch angle, $\theta \in (-\pi, \pi]$, where negative (positive) values indicate upstream (downstream) orientations.
Still, the mechanism of rheotaxis is not trivial.
Both pullers and pushers tend to swim upstream at weak flows, but they do so in a completely different fashion.

On the one hand, we describe the rheotaxis of pullers at low shear, as shown in the in the laboratory frame and the co-moving frame, respectively 
[see Videos 5, 6].
The three-sphere pullers tend to swim almost parallel to the surface, $\theta \lesssim -\pi/2$, with the director $\hat{\vec{t}}$ slightly pointing towards the surface.
Hence, the back sphere with the larger radius tends to stick out into the liquid where the flow gets stronger for larger $z$ values, so the puller can rotate against the flow.
This reorientation is referred to as the `weathervane effect', as described for example in Refs.~\cite{Hill2007, mathijssen2019oscillatory}.
The pullers tend to align with the shear plane rather slowly, taking tens to hundreds of oscillation periods.

On the other hand, we describe the pusher dynamics at low shear 
[see Videos 7, 8].
The three-sphere pushers tend to swim almost perpendicular to the surface, $\theta \gtrsim -\pi$, with the director $\hat{\vec{t}}$ slightly pointing upstream.
While the front sphere almost touches the surface, the back sphere sticks out into the flow so it gets advected downstream, leading to an upstream orientation.
Because the tail of the perpendicular pusher sticks out much further than the parallel puller, the `weathervane effect' is stronger, so the pushers have a much faster reorientation rate and only require a few three-sphere oscillations to turn upstream.
Rather than a burden, the cargo can therefore also be exploited to enhance rheotaxis.
This fundamental difference in the steady-state orientation also affects the velocity at which the two swimmer types can move against the flow. 
This is described in detail below, when we discuss the fixed point analysis.


\section{Swimming state diagrams}

Until now we have described the upstream motion at low shear, which is already fairly complex, but more intricate dynamics emerge at stronger flows.
We aim to quantify this systematically for different shear rates and initial conditions. 
Because the 3D dynamics reduce to 2D over time, we can cast them into a dynamical system where the relevant variables are the pitch angle, $\theta$, and the position of the central sphere, $z$. 
Figure~\ref{Fig2} shows the evolution of these dynamics in $(\theta, z)$ phase-space diagrams, where the top row shows the behaviours for pullers and the bottom row for pushers.
The steady-state swimming behaviours correspond to stable fixed points in these phase portraits, which change for different flow rates.

At weak flows, at $\gamDot = 1/3$, [Fig.~\ref{Fig2}(a)], the pullers mostly tend to swim upstream parallel to the surface (red), a stable fixed point around $(-{\pi}/{2}, {1}/{2})$, in agreement with the observations in Fig.~\ref{Fig1}. 
A small fraction of initial conditions also leads to downstream swimming parallel to the surface (blue), a stable fixed point around $({\pi}/{2}, {1}/{2})$.
The phase portraits corresponding to $\gamDot = 2/3$ and $\gamDot = 1$ are essentially the same as panel (a).
At strong flows, at $\gamDot = 2$, [Fig.~\ref{Fig2}(b)], almost all pullers are first advected downstream during a transient `toppling' motion.
However, over time they will end up in a stable state on the surface, oriented upstream.
If the external flow is stronger than the self-propulsion, this leads to downstream advection in the upstream orientation (green).
The transition of the final state from moving upstream (red) to downstream (green) occurs at $\gamDot \approx 1.33$, as discussed below.

The pushers show very different dynamics, because the two fixed points around $(\pm {\pi}/{2}, {1}/{2})$, of orientations parallel to the surface, are both unstable.
Instead, at $\gamDot = 1/3$, [Fig.~\ref{Fig2}(c)], the pushers tend to orient themselves almost normal to the wall (brown), but still a little directed upstream.
This corresponds to a fixed point around $(-0.8\pi, 0.9)$, which is  marked with an orange star.
Regardless of the initial conditions, all pushers end in this state, for all cases tested.
As the flow strength grows, the phase portraits corresponding to $\gamDot = 0.4$ and $\gamDot = 0.5$ remain essentially the same as panel (c).
At even larger shear rates, however, at $\gamDot = 1$,  [Fig.~\ref{Fig2}(d)], the orange star fixed point also becomes unstable, so the pushers tend to detach from the wall and topple downstream indefinitely (cyan).
These are the arc-like trajectories depicted in Fig.~\ref{Fig1}(f).

\section{Varying the flow strength}

\begin{figure*}[t]
	\includegraphics[width = 0.9 \linewidth]{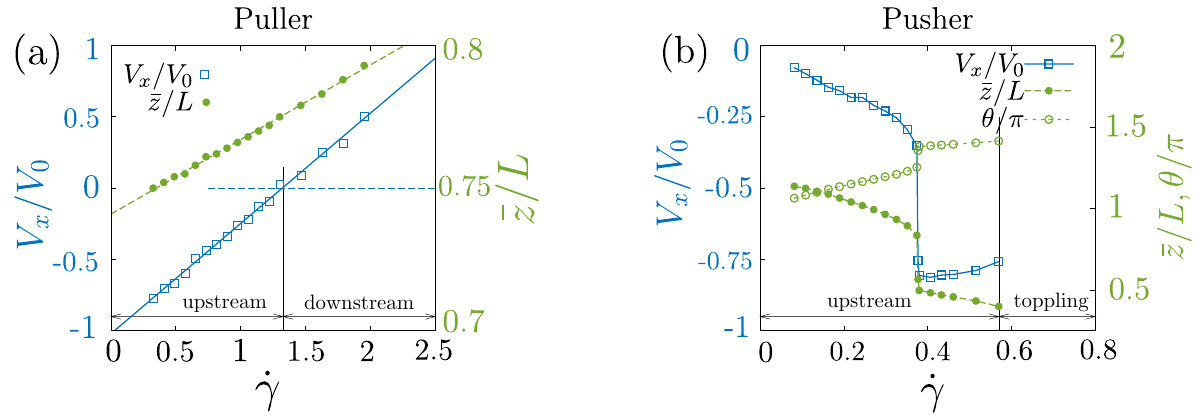}
	\caption{\label{Fig3}
	\textit{\textbf{Rheotactic performance}}. 
	We compare \textbf{(a)} pullers and \textbf{(b)} pushers, as a function of applied shear rate.
	Shown are the swimmer velocity $V_x(\gamDot)$ in blue squares, where negative values indicate upstream swimming, the pitch angle $\theta(\gamDot)$ in green open circles, and the position of the central sphere $z(\gamDot)$ in green filled circles. Note the different axes.
	Note, $V_0$ is the bulk speed of a neutral ($a_+=a$) swimmer in the absence of external flows and boundaries.
	}
\end{figure*}

Having identified the stable fixed points of the phase diagrams, we can determine the properties of these steady-state swimming modes as a function of shear rate.
In particular, we compute the velocity component $V_x(\gamDot)$, which is negative for upstream swimming,
the pitch angle $\theta(\gamDot)$, and the vertical position $z(\gamDot)$.
These quantities evolve very differently for pushers and pullers. 

Pullers in weak flows can move upstream very fast, $V_x \gtrsim - V_0$, almost their free swimming speed [Fig.~\ref{Fig3}(a)]. 
As the shear rate increases, $V_x$ increases linearly [blue line].
This trend is also enhanced because the vertical position gradually increases [green line], exposing the swimmer to more flow.
Therefore, the upstream swimming velocity tends to zero around $\gamDot_0 \approx 1.35$.
At higher shear the pullers are still oriented upstream, but they are advected downstream.

Surprisingly, the pushers show the opposite behaviour [Fig.~\ref{Fig3}(b)].
Their vertical position decreases with shear rate [filled circles], and the pitch angle changes from swimming perpendicular to parallel to the surface [open circles], so the swimmer is less exposed.
As a result, $V_x$ is almost zero in weak flows, but it decreases with shear rate, leading to faster upstream motion.
Moreover, around $\gamDot_c \approx 0.38$ there is a sharp transition.
The vertical position suddenly drops even further, so the upstream swimming speed also jumps up to $-V_x /V_0 \approx 0.8$.
At higher shear it stays relatively constant, until the pushers detach due to the toppling instability.
The critical shear rate at which this occurs is $\gamDot_t \approx 0.56$.

\section{Tuning the cargo size}

\begin{figure*}[t]
    \centering
    \includegraphics[width=0.9\linewidth]{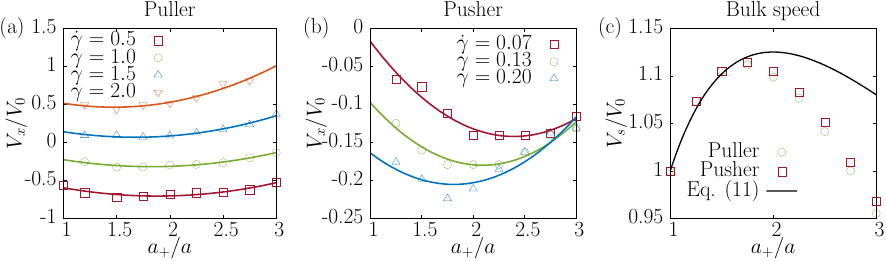}
    \caption{\label{Fig4}
	\textit{\textbf{Tuning the cargo size}}. 
	\textbf{(a)} The rheotactic performace of  pullers as a function of cargo size $a_+$, for different applied shear rates $\gamDot$. Shown is the swimmer velocity in the direction of the flow $V_x(\gamDot)$, where negative values indicate upstream swimming.
	As before, $V_0$ is the bulk speed of a neutral ($a_+=a$) swimmer, in the absence of external flows and boundaries.
	\textbf{(b)} The same for cargo pushers.
    \textbf{(c)} The bulk swimming speed as a function of cargo size, without flows and boundaries. We compare simulations for pullers (green circles) and pullers (red squared) with the linearised theory of Eq.~\ref{eq:SwimmingSpeed} (black line). 
	}
\end{figure*}

Next, we vary the cargo size across the range $a_+/a \in [1,3]$ both for pushers and pullers. 
Initially we confirmed that the mechanisms of rheotaxis remain unchanged. That is, the swimmers still reorient against the flow as described in Figures~\ref{Fig1}-\ref{Fig3}.
Then, when mapping out the upstream swimming velocity, we observe a non-trivial dependence on the cargo size [Fig.~\ref{Fig4}a,b]. 
With increasing $a_+$, both pushers and pullers first move faster against the flow, but for large cargo they move slower. So there is an optimal cargo size.

The position of this maximum can approximated analytically: 
Using the linearised theory of Eq.~(12-13) in Ref.~\cite{golestanian08} with $a_1 = a_2 = a$ and $a_3 = a_+$, arm length $L$ and oscillation amplitude $u_0$, we can approximate the bulk swimming speed as
    \begin{equation}
    \label{eq:SwimmingSpeed}
    V_s = \frac{21}{8} \frac{a^2 a_+}{(2a+a_+)^2} \frac{u_0^2}{L^2},
    \end{equation}
in the absence of flows and surfaces, as shown in Fig.~\ref{Fig4}c.
Equating to zero the derivative of Eq.~\eqref{eq:SwimmingSpeed} with respect to $a_+$, we find that this bulk swimming speed features a maximum at $a_+^*/a=2$.
However, the position of the maximum shifts because of the influence of the surface and the flow. 
This can be understood by considering the downstream advection speed $V_a = h \gamDot$, where $h=h(a_+)$ is the exposure height of the swimmer.
This expose increases with cargo size, so $h$ increases with $a_+$.
Therefore, the maximum of the upstream swimming speed $V_x \approx V_s \sin \theta + V_a$ shifts to smaller $a_+$ values for stronger flows, as expected.
Hence, the cargo size can be tuned to achieve the largest possible rheotactic performance.

\section{Discussion}

In summary, the rheotactic performance could be enhanced by exploiting the cargo, by tuning the swimmer geometry for a given shear rate.
Indeed, both cargo pushers and pullers tend to swim upstream near surfaces, but in a very different manner.
Pullers move almost parallel to the wall, so they are less susceptible to flow.
As a result, it takes longer to reorient against the flow, but their upstream swimming speed is generally large.
This speed decreases in strong currents, but even when detached they tend to return to the surface and move upstream.
Pushers, however, move almost perpendicular to the wall, so they are more susceptible to currents.
Consequently, they can reorient against the flow much faster, but their upstream swimming speed is poor at low shear.
Interestingly, this speed significantly improves at intermediate shear, to an extend that the pushers will actually outperform the pullers.
But in even stronger flows the pushers will detach from the wall and are washed downstream.
Thus, each cargo configuration has its own advantages, which may be optimised for different applications.
For example, if the swimmer were to be used to transport cargo \cite{golestanian08epje} upstream in fluctuating flow environments, it may be beneficial to use a puller for its robustness, while in strong but stable flows a pusher can be more expedient. 

A natural extension of our work would be to include effects of chirality, as observed in the dynamics of spermatozoa \cite{Miki2013, Kantsler2014} or bacterial flagella \cite{Kantsler2014}. 
This chirality induces an additional torque that leads to circular motion in the absence of flow \cite{Lauga2006}, but in flows it can lead to different dynamical regimes separated by critical shear rates \cite{mathijssen2019oscillatory}.
These predictions could be tested with three-sphere swimmers by introducing a counter-rotation to the head and tail spheres.
Furthermore, the effects of rotational diffusion, run-and-tumble dynamics and temporal motor variability \cite{figueroa2020spatial} would be required in the future to connect this work better with microbial ecology and bacterial contamination dynamics.
Or even more generally, besides the swimmer shape, also the swimming stroke may be tuned to design optimal navigation strategies.

Some important insights are revealed when comparing our work with related literature.
For autophoretic Janus (Au/Pt) nanorods, the pullers assume a larger tilt angle compared to pushers and they reorient faster against the flow \cite{brosseau2019relating}, while we see that the opposite is true for three-sphere swimmers.
For spherical squirmers \cite{uspal2015rheotaxis}, the pullers ($B_2/B_1>0$) also feature two stable fixed points facing upstream and downstream, like three-sphere swimmers, both almost parallel to the surface.
But unlike three-sphere swimmers, the majority of initial conditions leads to escape from the surface or downstream motion.
Spherical catalytic Janus particles can also move upstream near surfaces \cite{Palacci2015}.
Here a high surface coverage with catalyst results in orientations almost perpendicular to the wall, while a half coverage results in motion almost parallel to the wall \cite{uspal2015rheotaxis}.
These pitch angles may be observed in holography experiments \cite{bianchi2017holographic}.

This comparison with different types of micro-swimmers shows that the far-field hydrodynamic signature (dipole moment) is by itself not a good classifier of surface rheotaxis.
Instead, near-field flows must be considered for different systems, and the terms `puller' and `pusher' should rather be interpreted in terms of the swimmer shape itself. In our case that is whether the robot pushes or pulls cargo. 
However, the fact that different micro-robot types can employ a diversity of upstream swimming mechanisms need not be a disadvantage. 
If anything, it is interesting that there are different routes to the same goal.
It reemphasises our main result that rheotaxis can be regulated by swimmer shape and cargo size, which allows for tuning designs for specific applications. 

Indeed, throughout this paper we have studied the Najafi-Golestanian three-sphere swimmer because it is well established, analytically tractable, and easy to simulate to ensure reproducibility. 
An experimental realisation of this swimmer was developed recently \cite{grosjean2016realization}, made from a motile magnetocapillary self-assembly \cite{grosjean2018magnetocapillary}.
However, we expect that our conclusions will apply to a much wider class of micro-robots that push or pull cargo.  
Engineered or natural bacteria, for example, can resemble cargo pushers safe for the additional counter-rotation of the head and tail and run-tumble dynamics (see paragraph above). 
Dreyfus \textit{et al.} \cite{dreyfus2005microscopic} presented another related micro-robot that uses an active filament of magnetic beads to transport a red blood cell.
Also, phoretic colloids will likely be engineered with different shapes in future, to push or pull larger cargo vesicles. 
Moreover, recent advances in nanotechnology include the design of origami micro-machines \cite{miskin2018graphene, cui2019nanomagnetic}, four-dimensionally printed active materials \cite{ge2013active}, and artificial cilia \cite{den2008artificial, van2009printed, gu2020magnetic}. 
It would be very interesting if tuning the upstream swimming velocity of a micro-robot with cargo could be explored with these technologies.

\section{Acknowledgements}

We would like to thank Hartmut L\"{o}wen and Andreas M. Menzel for fruitful discussions and Jo Wenk for technical support. 
A.D.M.I. acknowledges support from the Deutsche Forschungsgemeinschaft (project DA~2107/1-1).
A.M. acknowledges funding from the Human Frontier Science Program (Fellowship LT001670/2017), and the United States Department of Agriculture (USDA-NIFA AFRI grant 2019-06706).


\section*{Supplementary information videos}

\begin{itemize}

		\item Video~1: Upstream swimming by cargo pullers. The external shear rate is weak, $\gamDot = 1/3$.
		Their motion is shown in the laboratory frame, projected on the $xy$ plane, as seen from above the surface.
		The swimmers are initially released from $z_0=1$ at various initial orientations $\phi_0$, all parallel to the surface, $\theta_0 = \pi/2$.
		All swimmers end up moving against the flow.

		\item Video~2:
		Upstream swimming by cargo pushers. The external shear rate is weak, $\gamDot = 1/3$.
		Their motion is shown in the laboratory frame, projected on the $xy$ plane, as seen from above the surface.
		The swimmers are initially released from $z_0=1$ at various initial orientations $\phi_0$, all parallel to the surface, $\theta_0 = \pi/2$.
		All swimmers end up moving against the flow.
	
		\item Video~3: Rheotaxis of a cargo puller in a strong flow, $\gamDot = 2$.
		Their motion is shown in the laboratory frame, projected on the $xy$ plane, as seen from above the surface.
		The swimmers are initially released from $z_0=1$ at various initial orientations $\phi_0$, all parallel to the surface, $\theta_0 = \pi/2$.
		All swimmers end up oriented upstream but they are advected downstream because the flow is too strong.
	
		\item Video~4: Rheotaxis of a cargo pusher in a strong flow, $\gamDot = 2$.
		Their motion is shown in the laboratory frame, projected on the $xy$ plane, as seen from above the surface.
		The swimmers are initially released from $z_0=1$ at various initial orientations $\phi_0$, all parallel to the surface, $\theta_0 = \pi/2$.
		All swimmers end up oriented upstream but they are advected downstream because the flow is too strong.

		\item Video~5:
		Upstream swimming by a cargo puller, shown in the laboratory frame. The external shear rate is $\gamDot = 2/3$. 
		The swimmer position in 3D space is shown in blue, its projection onto the $xz$ plane is shown in orange, and its projection onto the $yz$ plane is shown in green. This reveals how its orientation evolves over time.
		The swimmer is initially released from $z_0=1$ at initial orientations $\phi_0 = \pi/2$ and parallel to the surface, $\theta_0 = \pi/2$.
		The swimmer ends up moving against the flow.
	
		\item Video~6:
		Same as 
		Video~5,
		shown in the frame co-moving with the swimmer.
	
		\item Video~7:
		Upstream swimming by a cargo pusher, shown in the laboratory frame. The external shear rate is $\gamDot = 1/3$. 
		The swimmer position in 3D space is shown in blue, its projection onto the $xz$ plane is shown in orange, and its projection onto the $yz$ plane is shown in green. This reveals how its orientation evolves over time.
		The swimmer is initially released from $z_0=1$ at initial orientations $\phi_0 = \pi/2$ and parallel to the surface, $\theta_0 = \pi/2$.
		The swimmer ends up moving against the flow.
	
		\item Video~8:
		Same as 
		Video~7,
		shown in the frame co-moving with the swimmer.
	
\end{itemize}


\input{main.bbl}

\end{document}

%% file: commands.tex
\newcommand{\vect}[1]{\bm{#1}}

\newcommand{\R}{\vect{r}}
\newcommand{\Intd}{\mathrm{d }}

\newcommand{\hatez}{\vect{\hat{e}}_{z}}
\newcommand{\hatex}{\vect{\hat{e}}_{x}}
\newcommand{\hatey}{\vect{\hat{e}}_{y}}

\newcommand{\bNabla}{\boldsymbol{\nabla}}

\newcommand{\gamDot}{\dot{\gamma}}

\newcommand{\bomega}{\boldsymbol{\omega}}
\newcommand{\bOmega}{\boldsymbol{\Omega}}

\newcommand{\bmu}{\boldsymbol{\mu}}

\newcommand{\hatt}{\vect{\hat{t}}}
\newcommand{\hattheta}{\vect{\hat{\theta}}}
\newcommand{\hatphi}{\vect{\hat{\phi}}}

%% file: main.bbl
%

%% file: main.bbl
\begin{thebibliography}{67}%
\makeatletter
\providecommand \@ifxundefined [1]{%
 \@ifx{#1\undefined}
}%
\providecommand \@ifnum [1]{%
 \ifnum #1\expandafter \@firstoftwo
 \else \expandafter \@secondoftwo
 \fi
}%
\providecommand \@ifx [1]{%
 \ifx #1\expandafter \@firstoftwo
 \else \expandafter \@secondoftwo
 \fi
}%
\providecommand \natexlab [1]{#1}%
\providecommand \enquote  [1]{``#1''}%
\providecommand \bibnamefont  [1]{#1}%
\providecommand \bibfnamefont [1]{#1}%
\providecommand \citenamefont [1]{#1}%
\providecommand \href@noop [0]{\@secondoftwo}%
\providecommand \href [0]{\begingroup \@sanitize@url \@href}%
\providecommand \@href[1]{\@@startlink{#1}\@@href}%
\providecommand \@@href[1]{\endgroup#1\@@endlink}%
\providecommand \@sanitize@url [0]{\catcode `\\12\catcode `\$12\catcode
  `\&12\catcode `\#12\catcode `\^12\catcode `\_12\catcode `\%12\relax}%
\providecommand \@@startlink[1]{}%
\providecommand \@@endlink[0]{}%
\providecommand \url  [0]{\begingroup\@sanitize@url \@url }%
\providecommand \@url [1]{\endgroup\@href {#1}{\urlprefix }}%
\providecommand \urlprefix  [0]{URL }%
\providecommand \Eprint [0]{\href }%
\providecommand \doibase [0]{https://doi.org/}%
\providecommand \selectlanguage [0]{\@gobble}%
\providecommand \bibinfo  [0]{\@secondoftwo}%
\providecommand \bibfield  [0]{\@secondoftwo}%
\providecommand \translation [1]{[#1]}%
\providecommand \BibitemOpen [0]{}%
\providecommand \bibitemStop [0]{}%
\providecommand \bibitemNoStop [0]{.\EOS\space}%
\providecommand \EOS [0]{\spacefactor3000\relax}%
\providecommand \BibitemShut  [1]{\csname bibitem#1\endcsname}%
\let\auto@bib@innerbib\@empty
\bibitem [{\citenamefont {Purcell}(1977)}]{purcell1977}%
  \BibitemOpen
  \bibfield  {author} {\bibinfo {author} {\bibfnamefont {E.~M.}\ \bibnamefont
  {Purcell}},\ }\href {https://doi.org/10.1119/1.10903} {\bibfield  {journal}
  {\bibinfo  {journal} {Amer. J. Phys.}\ }\textbf {\bibinfo {volume} {45}},\
  \bibinfo {pages} {3} (\bibinfo {year} {1977})}\BibitemShut {NoStop}%
\bibitem [{\citenamefont {Lauga}\ and\ \citenamefont
  {Powers}(2009)}]{Lauga2009}%
  \BibitemOpen
  \bibfield  {author} {\bibinfo {author} {\bibfnamefont {E.}~\bibnamefont
  {Lauga}}\ and\ \bibinfo {author} {\bibfnamefont {T.~R.}\ \bibnamefont
  {Powers}},\ }\href {https://doi.org/10.1088/0034-4885/72/9/096601} {\bibfield
   {journal} {\bibinfo  {journal} {Rep. Prog. Phys.}\ }\textbf {\bibinfo
  {volume} {72}},\ \bibinfo {pages} {096601} (\bibinfo {year}
  {2009})}\BibitemShut {NoStop}%
\bibitem [{\citenamefont {Koch}\ and\ \citenamefont
  {Subramanian}(2011)}]{koch2011collective}%
  \BibitemOpen
  \bibfield  {author} {\bibinfo {author} {\bibfnamefont {D.~L.}\ \bibnamefont
  {Koch}}\ and\ \bibinfo {author} {\bibfnamefont {G.}~\bibnamefont
  {Subramanian}},\ }\href {https://doi.org/10.1146/annurev-fluid-121108-145434}
  {\bibfield  {journal} {\bibinfo  {journal} {Ann. Rev. Fluid Mech.}\ }\textbf
  {\bibinfo {volume} {43}},\ \bibinfo {pages} {637} (\bibinfo {year}
  {2011})}\BibitemShut {NoStop}%
\bibitem [{\citenamefont {Elgeti}\ \emph {et~al.}(2015)\citenamefont {Elgeti},
  \citenamefont {Winkler},\ and\ \citenamefont {Gompper}}]{elgeti2015}%
  \BibitemOpen
  \bibfield  {author} {\bibinfo {author} {\bibfnamefont {J.}~\bibnamefont
  {Elgeti}}, \bibinfo {author} {\bibfnamefont {R.~G.}\ \bibnamefont
  {Winkler}},\ and\ \bibinfo {author} {\bibfnamefont {G.}~\bibnamefont
  {Gompper}},\ }\href {https://doi.org/10.1088/0034-4885/78/5/056601}
  {\bibfield  {journal} {\bibinfo  {journal} {Rep. Progr. Phys.}\ }\textbf
  {\bibinfo {volume} {78}},\ \bibinfo {pages} {056601} (\bibinfo {year}
  {2015})}\BibitemShut {NoStop}%
\bibitem [{\citenamefont {Bechinger}\ \emph {et~al.}(2016)\citenamefont
  {Bechinger}, \citenamefont {Di~Leonardo}, \citenamefont {L{\"o}wen},
  \citenamefont {Reichhardt}, \citenamefont {Volpe},\ and\ \citenamefont
  {Volpe}}]{bechinger2016active}%
  \BibitemOpen
  \bibfield  {author} {\bibinfo {author} {\bibfnamefont {C.}~\bibnamefont
  {Bechinger}}, \bibinfo {author} {\bibfnamefont {R.}~\bibnamefont
  {Di~Leonardo}}, \bibinfo {author} {\bibfnamefont {H.}~\bibnamefont
  {L{\"o}wen}}, \bibinfo {author} {\bibfnamefont {C.}~\bibnamefont
  {Reichhardt}}, \bibinfo {author} {\bibfnamefont {G.}~\bibnamefont {Volpe}},\
  and\ \bibinfo {author} {\bibfnamefont {G.}~\bibnamefont {Volpe}},\ }\href
  {https://doi.org/10.1103/RevModPhys.88.045006} {\bibfield  {journal}
  {\bibinfo  {journal} {Rev. Mod. Phys.}\ }\textbf {\bibinfo {volume} {88}},\
  \bibinfo {pages} {045006} (\bibinfo {year} {2016})}\BibitemShut {NoStop}%
\bibitem [{\citenamefont {Wheeler}\ \emph {et~al.}(2019)\citenamefont
  {Wheeler}, \citenamefont {Secchi}, \citenamefont {Rusconi},\ and\
  \citenamefont {Stocker}}]{wheeler2019not}%
  \BibitemOpen
  \bibfield  {author} {\bibinfo {author} {\bibfnamefont {J.~D.}\ \bibnamefont
  {Wheeler}}, \bibinfo {author} {\bibfnamefont {E.}~\bibnamefont {Secchi}},
  \bibinfo {author} {\bibfnamefont {R.}~\bibnamefont {Rusconi}},\ and\ \bibinfo
  {author} {\bibfnamefont {R.}~\bibnamefont {Stocker}},\ }\href
  {https://doi.org/10.1146/annurev-cellbio-100818-125119} {\bibfield  {journal}
  {\bibinfo  {journal} {Ann. Rev. Cell. Develop. Biol.}\ }\textbf {\bibinfo
  {volume} {35}},\ \bibinfo {pages} {213} (\bibinfo {year} {2019})}\BibitemShut
  {NoStop}%
\bibitem [{\citenamefont {Rusconi}\ and\ \citenamefont
  {Stocker}(2015)}]{rusconi2015}%
  \BibitemOpen
  \bibfield  {author} {\bibinfo {author} {\bibfnamefont {R.}~\bibnamefont
  {Rusconi}}\ and\ \bibinfo {author} {\bibfnamefont {R.}~\bibnamefont
  {Stocker}},\ }\href
  {https://doi.org/https://doi.org/10.1016/j.mib.2015.03.003} {\bibfield
  {journal} {\bibinfo  {journal} {Curr. Opin. Microbiol.}\ }\textbf {\bibinfo
  {volume} {25}},\ \bibinfo {pages} {1 } (\bibinfo {year} {2015})}\BibitemShut
  {NoStop}%
\bibitem [{\citenamefont {de~Graaf}\ \emph
  {et~al.}(2016{\natexlab{a}})\citenamefont {de~Graaf}, \citenamefont {Menke},
  \citenamefont {Mathijssen}, \citenamefont {Fabritius}, \citenamefont {Holm},\
  and\ \citenamefont {Shendruk}}]{degraaf16}%
  \BibitemOpen
  \bibfield  {author} {\bibinfo {author} {\bibfnamefont {J.}~\bibnamefont
  {de~Graaf}}, \bibinfo {author} {\bibfnamefont {H.}~\bibnamefont {Menke}},
  \bibinfo {author} {\bibfnamefont {A.~J. T.~M.}\ \bibnamefont {Mathijssen}},
  \bibinfo {author} {\bibfnamefont {M.}~\bibnamefont {Fabritius}}, \bibinfo
  {author} {\bibfnamefont {C.}~\bibnamefont {Holm}},\ and\ \bibinfo {author}
  {\bibfnamefont {T.~N.}\ \bibnamefont {Shendruk}},\ }\href
  {https://doi.org/10.1063/1.4944962} {\bibfield  {journal} {\bibinfo
  {journal} {J. Chem. Phys.}\ }\textbf {\bibinfo {volume} {144}},\ \bibinfo
  {pages} {134106} (\bibinfo {year} {2016}{\natexlab{a}})}\BibitemShut
  {NoStop}%
\bibitem [{\citenamefont {Berke}\ \emph {et~al.}(2008)\citenamefont {Berke},
  \citenamefont {Turner}, \citenamefont {Berg},\ and\ \citenamefont
  {Lauga}}]{berke08}%
  \BibitemOpen
  \bibfield  {author} {\bibinfo {author} {\bibfnamefont {A.~P.}\ \bibnamefont
  {Berke}}, \bibinfo {author} {\bibfnamefont {L.}~\bibnamefont {Turner}},
  \bibinfo {author} {\bibfnamefont {H.~C.}\ \bibnamefont {Berg}},\ and\
  \bibinfo {author} {\bibfnamefont {E.}~\bibnamefont {Lauga}},\ }\href
  {https://doi.org/10.1103/PhysRevLett.101.038102} {\bibfield  {journal}
  {\bibinfo  {journal} {Phys. Rev. Lett.}\ }\textbf {\bibinfo {volume} {101}},\
  \bibinfo {pages} {038102} (\bibinfo {year} {2008})}\BibitemShut {NoStop}%
\bibitem [{\citenamefont {Lauga}\ \emph {et~al.}(2006)\citenamefont {Lauga},
  \citenamefont {DiLuzio}, \citenamefont {Whitesides},\ and\ \citenamefont
  {Stone}}]{Lauga2006}%
  \BibitemOpen
  \bibfield  {author} {\bibinfo {author} {\bibfnamefont {E.}~\bibnamefont
  {Lauga}}, \bibinfo {author} {\bibfnamefont {W.~R.}\ \bibnamefont {DiLuzio}},
  \bibinfo {author} {\bibfnamefont {G.~M.}\ \bibnamefont {Whitesides}},\ and\
  \bibinfo {author} {\bibfnamefont {H.~A.}\ \bibnamefont {Stone}},\ }\href
  {https://doi.org/10.1529/biophysj.105.069401} {\bibfield  {journal} {\bibinfo
   {journal} {Biophys. J.}\ }\textbf {\bibinfo {volume} {90}},\ \bibinfo
  {pages} {400} (\bibinfo {year} {2006})}\BibitemShut {NoStop}%
\bibitem [{\citenamefont {Spagnolie}\ and\ \citenamefont
  {Lauga}(2012)}]{spagnolie12}%
  \BibitemOpen
  \bibfield  {author} {\bibinfo {author} {\bibfnamefont {S.~E.}\ \bibnamefont
  {Spagnolie}}\ and\ \bibinfo {author} {\bibfnamefont {E.}~\bibnamefont
  {Lauga}},\ }\href {https://doi.org/10.1017/jfm.2012.101} {\bibfield
  {journal} {\bibinfo  {journal} {J. Fluid Mech.}\ }\textbf {\bibinfo {volume}
  {700}},\ \bibinfo {pages} {105} (\bibinfo {year} {2012})}\BibitemShut
  {NoStop}%
\bibitem [{\citenamefont {Miño}\ \emph {et~al.}(2018)\citenamefont {Miño},
  \citenamefont {Baabour}, \citenamefont {Chertcoff}, \citenamefont {Gutkind},
  \citenamefont {Clément}, \citenamefont {Auradou},\ and\ \citenamefont
  {Ippolito}}]{mino2018}%
  \BibitemOpen
  \bibfield  {author} {\bibinfo {author} {\bibfnamefont {G.}~\bibnamefont
  {Miño}}, \bibinfo {author} {\bibfnamefont {M.}~\bibnamefont {Baabour}},
  \bibinfo {author} {\bibfnamefont {R.}~\bibnamefont {Chertcoff}}, \bibinfo
  {author} {\bibfnamefont {G.}~\bibnamefont {Gutkind}}, \bibinfo {author}
  {\bibfnamefont {E.}~\bibnamefont {Clément}}, \bibinfo {author}
  {\bibfnamefont {H.}~\bibnamefont {Auradou}},\ and\ \bibinfo {author}
  {\bibfnamefont {I.}~\bibnamefont {Ippolito}},\ }\href
  {https://doi.org/10.4236/aim.2018.86030} {\bibfield  {journal} {\bibinfo
  {journal} {Adv. Microbiol.}\ }\textbf {\bibinfo {volume} {8}},\ \bibinfo
  {pages} {451} (\bibinfo {year} {2018})}\BibitemShut {NoStop}%
\bibitem [{\citenamefont {Kessler}(1985)}]{Kessler1985}%
  \BibitemOpen
  \bibfield  {author} {\bibinfo {author} {\bibfnamefont {J.~O.}\ \bibnamefont
  {Kessler}},\ }\href {https://doi.org/10.1038/3132} {\bibfield  {journal}
  {\bibinfo  {journal} {Nature}\ }\textbf {\bibinfo {volume} {313}},\ \bibinfo
  {pages} {218} (\bibinfo {year} {1985})}\BibitemShut {NoStop}%
\bibitem [{\citenamefont {Rusconi}\ \emph {et~al.}(2014)\citenamefont
  {Rusconi}, \citenamefont {Guasto},\ and\ \citenamefont
  {Stocker}}]{Rusconi2014}%
  \BibitemOpen
  \bibfield  {author} {\bibinfo {author} {\bibfnamefont {R.}~\bibnamefont
  {Rusconi}}, \bibinfo {author} {\bibfnamefont {J.}~\bibnamefont {Guasto}},\
  and\ \bibinfo {author} {\bibfnamefont {R.}~\bibnamefont {Stocker}},\ }\href
  {https://doi.org/10.1038/NPHYS2883} {\bibfield  {journal} {\bibinfo
  {journal} {Nat. Phys.}\ }\textbf {\bibinfo {volume} {10}},\ \bibinfo {pages}
  {212} (\bibinfo {year} {2014})}\BibitemShut {NoStop}%
\bibitem [{\citenamefont {Barry}\ \emph {et~al.}(2015)\citenamefont {Barry},
  \citenamefont {Rusconi}, \citenamefont {Guasto},\ and\ \citenamefont
  {Stocker}}]{Barry2015}%
  \BibitemOpen
  \bibfield  {author} {\bibinfo {author} {\bibfnamefont {M.~T.}\ \bibnamefont
  {Barry}}, \bibinfo {author} {\bibfnamefont {R.}~\bibnamefont {Rusconi}},
  \bibinfo {author} {\bibfnamefont {J.~S.}\ \bibnamefont {Guasto}},\ and\
  \bibinfo {author} {\bibfnamefont {R.}~\bibnamefont {Stocker}},\ }\href
  {https://doi.org/10.1098/rsif.2015.0791} {\bibfield  {journal} {\bibinfo
  {journal} {J. R. Soc. Interface}\ }\textbf {\bibinfo {volume} {12}},\
  \bibinfo {pages} {20150791} (\bibinfo {year} {2015})}\BibitemShut {NoStop}%
\bibitem [{\citenamefont {DiLuzio}\ \emph {et~al.}(2005)\citenamefont
  {DiLuzio}, \citenamefont {Turner}, \citenamefont {Mayer}, \citenamefont
  {Garstecki}, \citenamefont {Weibel}, \citenamefont {Berg},\ and\
  \citenamefont {Whitesides}}]{Luzio2005}%
  \BibitemOpen
  \bibfield  {author} {\bibinfo {author} {\bibfnamefont {W.~R.}\ \bibnamefont
  {DiLuzio}}, \bibinfo {author} {\bibfnamefont {L.}~\bibnamefont {Turner}},
  \bibinfo {author} {\bibfnamefont {M.}~\bibnamefont {Mayer}}, \bibinfo
  {author} {\bibfnamefont {P.}~\bibnamefont {Garstecki}}, \bibinfo {author}
  {\bibfnamefont {D.~B.}\ \bibnamefont {Weibel}}, \bibinfo {author}
  {\bibfnamefont {H.~C.}\ \bibnamefont {Berg}},\ and\ \bibinfo {author}
  {\bibfnamefont {G.~M.}\ \bibnamefont {Whitesides}},\ }\href
  {https://doi.org/10.1038/nature03660} {\bibfield  {journal} {\bibinfo
  {journal} {Nature}\ }\textbf {\bibinfo {volume} {435}},\ \bibinfo {pages}
  {1271} (\bibinfo {year} {2005})}\BibitemShut {NoStop}%
\bibitem [{\citenamefont {de~Graaf}\ \emph
  {et~al.}(2016{\natexlab{b}})\citenamefont {de~Graaf}, \citenamefont
  {Mathijssen}, \citenamefont {Fabritius}, \citenamefont {Menke}, \citenamefont
  {Holm},\ and\ \citenamefont {Shendruk}}]{DeGraaf2016}%
  \BibitemOpen
  \bibfield  {author} {\bibinfo {author} {\bibfnamefont {J.}~\bibnamefont
  {de~Graaf}}, \bibinfo {author} {\bibfnamefont {A.~J.}\ \bibnamefont
  {Mathijssen}}, \bibinfo {author} {\bibfnamefont {M.}~\bibnamefont
  {Fabritius}}, \bibinfo {author} {\bibfnamefont {H.}~\bibnamefont {Menke}},
  \bibinfo {author} {\bibfnamefont {C.}~\bibnamefont {Holm}},\ and\ \bibinfo
  {author} {\bibfnamefont {T.~N.}\ \bibnamefont {Shendruk}},\ }\href
  {https://doi.org/10.1039/C6SM00939E} {\bibfield  {journal} {\bibinfo
  {journal} {Soft Matter}\ }\textbf {\bibinfo {volume} {12}},\ \bibinfo {pages}
  {4704} (\bibinfo {year} {2016}{\natexlab{b}})}\BibitemShut {NoStop}%
\bibitem [{\citenamefont {Maldonado}\ and\ \citenamefont
  {Latz}(2007)}]{maldonado2007}%
  \BibitemOpen
  \bibfield  {author} {\bibinfo {author} {\bibfnamefont {E.~M.}\ \bibnamefont
  {Maldonado}}\ and\ \bibinfo {author} {\bibfnamefont {M.~I.}\ \bibnamefont
  {Latz}},\ }\href {https://doi.org/10.2307/25066606} {\bibfield  {journal}
  {\bibinfo  {journal} {Biol. Bull.}\ }\textbf {\bibinfo {volume} {212}},\
  \bibinfo {pages} {242} (\bibinfo {year} {2007})},\ \bibinfo {note} {pMID:
  17565113}\BibitemShut {NoStop}%
\bibitem [{\citenamefont {Mathijssen}\ \emph
  {et~al.}(2019{\natexlab{a}})\citenamefont {Mathijssen}, \citenamefont
  {Culver}, \citenamefont {Bhamla},\ and\ \citenamefont
  {Prakash}}]{mathijssen2019collective}%
  \BibitemOpen
  \bibfield  {author} {\bibinfo {author} {\bibfnamefont {A.~J. T.~M.}\
  \bibnamefont {Mathijssen}}, \bibinfo {author} {\bibfnamefont
  {J.}~\bibnamefont {Culver}}, \bibinfo {author} {\bibfnamefont {M.~S.}\
  \bibnamefont {Bhamla}},\ and\ \bibinfo {author} {\bibfnamefont
  {M.}~\bibnamefont {Prakash}},\ }\href
  {https://doi.org/10.1038/s41586-019-1387-9} {\bibfield  {journal} {\bibinfo
  {journal} {Nature}\ }\textbf {\bibinfo {volume} {571}},\ \bibinfo {pages}
  {560} (\bibinfo {year} {2019}{\natexlab{a}})}\BibitemShut {NoStop}%
\bibitem [{\citenamefont {Chengala}\ \emph {et~al.}(2013)\citenamefont
  {Chengala}, \citenamefont {Hondzo},\ and\ \citenamefont
  {Sheng}}]{chengala2013}%
  \BibitemOpen
  \bibfield  {author} {\bibinfo {author} {\bibfnamefont {A.}~\bibnamefont
  {Chengala}}, \bibinfo {author} {\bibfnamefont {M.}~\bibnamefont {Hondzo}},\
  and\ \bibinfo {author} {\bibfnamefont {J.}~\bibnamefont {Sheng}},\ }\href
  {https://doi.org/10.1103/PhysRevE.87.052704} {\bibfield  {journal} {\bibinfo
  {journal} {Phys. Rev. E}\ }\textbf {\bibinfo {volume} {87}},\ \bibinfo
  {pages} {052704} (\bibinfo {year} {2013})}\BibitemShut {NoStop}%
\bibitem [{\citenamefont {Jeffery}(1922)}]{Jeffery1922}%
  \BibitemOpen
  \bibfield  {author} {\bibinfo {author} {\bibfnamefont {G.~B.}\ \bibnamefont
  {Jeffery}},\ }\href {https://doi.org/10.1098/rspa.1922.0078} {\bibfield
  {journal} {\bibinfo  {journal} {Proc. Roy. Soc. A}\ }\textbf {\bibinfo
  {volume} {102}},\ \bibinfo {pages} {161} (\bibinfo {year}
  {1922})}\BibitemShut {NoStop}%
\bibitem [{\citenamefont {Z{\"{o}}ttl}\ and\ \citenamefont
  {Stark}(2012)}]{Zoettl2012}%
  \BibitemOpen
  \bibfield  {author} {\bibinfo {author} {\bibfnamefont {A.}~\bibnamefont
  {Z{\"{o}}ttl}}\ and\ \bibinfo {author} {\bibfnamefont {H.}~\bibnamefont
  {Stark}},\ }\href {https://doi.org/10.1103/PhysRevLett.108.218104} {\bibfield
   {journal} {\bibinfo  {journal} {Phys. Rev. Lett.}\ }\textbf {\bibinfo
  {volume} {108}},\ \bibinfo {pages} {218104} (\bibinfo {year}
  {2012})}\BibitemShut {NoStop}%
\bibitem [{\citenamefont {Junot}\ \emph {et~al.}(2019)\citenamefont {Junot},
  \citenamefont {Figueroa-Morales}, \citenamefont {Darnige}, \citenamefont
  {Lindner}, \citenamefont {Soto}, \citenamefont {Auradou},\ and\ \citenamefont
  {Cl{\'{e}}ment}}]{junot2019bacterium}%
  \BibitemOpen
  \bibfield  {author} {\bibinfo {author} {\bibfnamefont {G.}~\bibnamefont
  {Junot}}, \bibinfo {author} {\bibfnamefont {N.}~\bibnamefont
  {Figueroa-Morales}}, \bibinfo {author} {\bibfnamefont {T.}~\bibnamefont
  {Darnige}}, \bibinfo {author} {\bibfnamefont {A.}~\bibnamefont {Lindner}},
  \bibinfo {author} {\bibfnamefont {R.}~\bibnamefont {Soto}}, \bibinfo {author}
  {\bibfnamefont {H.}~\bibnamefont {Auradou}},\ and\ \bibinfo {author}
  {\bibfnamefont {E.}~\bibnamefont {Cl{\'{e}}ment}},\ }\href
  {https://doi.org/10.1209/0295-5075/126/44003} {\bibfield  {journal} {\bibinfo
   {journal} {{EPL} (Europhysics Letters)}\ }\textbf {\bibinfo {volume}
  {126}},\ \bibinfo {pages} {44003} (\bibinfo {year} {2019})}\BibitemShut
  {NoStop}%
\bibitem [{\citenamefont {Marcos}\ \emph {et~al.}(2009)\citenamefont {Marcos},
  \citenamefont {Fu}, \citenamefont {Powers},\ and\ \citenamefont
  {Stocker}}]{marcos2009separation}%
  \BibitemOpen
  \bibfield  {author} {\bibinfo {author} {\bibnamefont {Marcos}}, \bibinfo
  {author} {\bibfnamefont {H.~C.}\ \bibnamefont {Fu}}, \bibinfo {author}
  {\bibfnamefont {T.~R.}\ \bibnamefont {Powers}},\ and\ \bibinfo {author}
  {\bibfnamefont {R.}~\bibnamefont {Stocker}},\ }\href
  {https://doi.org/10.1103/PhysRevLett.102.158103} {\bibfield  {journal}
  {\bibinfo  {journal} {Phys. Rev. Lett.}\ }\textbf {\bibinfo {volume} {102}},\
  \bibinfo {pages} {158103} (\bibinfo {year} {2009})}\BibitemShut {NoStop}%
\bibitem [{\citenamefont {Marcos}\ \emph {et~al.}(2012)\citenamefont {Marcos},
  \citenamefont {Fu}, \citenamefont {Powers},\ and\ \citenamefont
  {Stocker}}]{Marcos2012}%
  \BibitemOpen
  \bibfield  {author} {\bibinfo {author} {\bibnamefont {Marcos}}, \bibinfo
  {author} {\bibfnamefont {H.~C.}\ \bibnamefont {Fu}}, \bibinfo {author}
  {\bibfnamefont {T.~R.}\ \bibnamefont {Powers}},\ and\ \bibinfo {author}
  {\bibfnamefont {R.}~\bibnamefont {Stocker}},\ }\href
  {https://doi.org/10.1073/pnas.1120955109} {\bibfield  {journal} {\bibinfo
  {journal} {Proc. Nat. Acad. Sci.}\ }\textbf {\bibinfo {volume} {109}},\
  \bibinfo {pages} {4780} (\bibinfo {year} {2012})}\BibitemShut {NoStop}%
\bibitem [{\citenamefont {Jing}\ \emph {et~al.}(2020)\citenamefont {Jing},
  \citenamefont {Z{\"o}ttl}, \citenamefont {Cl{\'e}ment},\ and\ \citenamefont
  {Lindner}}]{jing2020chirality}%
  \BibitemOpen
  \bibfield  {author} {\bibinfo {author} {\bibfnamefont {G.}~\bibnamefont
  {Jing}}, \bibinfo {author} {\bibfnamefont {A.}~\bibnamefont {Z{\"o}ttl}},
  \bibinfo {author} {\bibfnamefont {{\'E}.}~\bibnamefont {Cl{\'e}ment}},\ and\
  \bibinfo {author} {\bibfnamefont {A.}~\bibnamefont {Lindner}},\ }\href
  {https://arxiv.org/abs/2003.04012} {\bibfield  {journal} {\bibinfo  {journal}
  {arXiv preprint}\ ,\ \bibinfo {pages} {2003.04012}} (\bibinfo {year}
  {2020})}\BibitemShut {NoStop}%
\bibitem [{\citenamefont {Mathijssen}\ \emph {et~al.}(2016)\citenamefont
  {Mathijssen}, \citenamefont {Shendruk}, \citenamefont {Yeomans},\ and\
  \citenamefont {Doostmohammadi}}]{Mathijssen2016}%
  \BibitemOpen
  \bibfield  {author} {\bibinfo {author} {\bibfnamefont {A.~J. T.~M.}\
  \bibnamefont {Mathijssen}}, \bibinfo {author} {\bibfnamefont {T.~N.}\
  \bibnamefont {Shendruk}}, \bibinfo {author} {\bibfnamefont {J.~M.}\
  \bibnamefont {Yeomans}},\ and\ \bibinfo {author} {\bibfnamefont
  {A.}~\bibnamefont {Doostmohammadi}},\ }\href
  {https://doi.org/10.1103/PhysRevLett.116.028104} {\bibfield  {journal}
  {\bibinfo  {journal} {Phys. Rev. Lett.}\ }\textbf {\bibinfo {volume} {116}},\
  \bibinfo {pages} {028104} (\bibinfo {year} {2016})}\BibitemShut {NoStop}%
\bibitem [{\citenamefont {Bretherton}(1962)}]{bretherton62}%
  \BibitemOpen
  \bibfield  {author} {\bibinfo {author} {\bibfnamefont {F.~P.}\ \bibnamefont
  {Bretherton}},\ }\href {https://doi.org/10.1017/S002211206200124X} {\bibfield
   {journal} {\bibinfo  {journal} {J. Fluid Mech.}\ }\textbf {\bibinfo {volume}
  {14}},\ \bibinfo {pages} {284} (\bibinfo {year} {1962})}\BibitemShut
  {NoStop}%
\bibitem [{\citenamefont {Miki}\ and\ \citenamefont
  {Clapham}(2013)}]{Miki2013}%
  \BibitemOpen
  \bibfield  {author} {\bibinfo {author} {\bibfnamefont {K.}~\bibnamefont
  {Miki}}\ and\ \bibinfo {author} {\bibfnamefont {D.~E.}\ \bibnamefont
  {Clapham}},\ }\href {https://doi.org/10.1016/j.cub.2013.02.007} {\bibfield
  {journal} {\bibinfo  {journal} {Curr. Biol.}\ }\textbf {\bibinfo {volume}
  {23}},\ \bibinfo {pages} {443} (\bibinfo {year} {2013})}\BibitemShut
  {NoStop}%
\bibitem [{\citenamefont {Kantsler}\ \emph {et~al.}(2014)\citenamefont
  {Kantsler}, \citenamefont {Dunkel}, \citenamefont {Blayney},\ and\
  \citenamefont {Goldstein}}]{Kantsler2014}%
  \BibitemOpen
  \bibfield  {author} {\bibinfo {author} {\bibfnamefont {V.}~\bibnamefont
  {Kantsler}}, \bibinfo {author} {\bibfnamefont {J.}~\bibnamefont {Dunkel}},
  \bibinfo {author} {\bibfnamefont {M.}~\bibnamefont {Blayney}},\ and\ \bibinfo
  {author} {\bibfnamefont {R.~E.}\ \bibnamefont {Goldstein}},\ }\href
  {https://doi.org/10.7554/eLife.02403} {\bibfield  {journal} {\bibinfo
  {journal} {e{L}ife}\ ,\ \bibinfo {pages} {02403}} (\bibinfo {year}
  {2014})}\BibitemShut {NoStop}%
\bibitem [{\citenamefont {Tung}\ \emph {et~al.}(2015)\citenamefont {Tung},
  \citenamefont {Ardon}, \citenamefont {Roy}, \citenamefont {Koch},
  \citenamefont {Suarez},\ and\ \citenamefont {Wu}}]{Tung2015}%
  \BibitemOpen
  \bibfield  {author} {\bibinfo {author} {\bibfnamefont {C.~K.}\ \bibnamefont
  {Tung}}, \bibinfo {author} {\bibfnamefont {F.}~\bibnamefont {Ardon}},
  \bibinfo {author} {\bibfnamefont {A.}~\bibnamefont {Roy}}, \bibinfo {author}
  {\bibfnamefont {D.~L.}\ \bibnamefont {Koch}}, \bibinfo {author}
  {\bibfnamefont {S.~S.}\ \bibnamefont {Suarez}},\ and\ \bibinfo {author}
  {\bibfnamefont {M.}~\bibnamefont {Wu}},\ }\href
  {https://doi.org/10.1103/PhysRevLett.114.108102} {\bibfield  {journal}
  {\bibinfo  {journal} {Phys. Rev. Lett.}\ }\textbf {\bibinfo {volume} {114}},\
  \bibinfo {pages} {108102} (\bibinfo {year} {2015})}\BibitemShut {NoStop}%
\bibitem [{\citenamefont {Hill}\ \emph {et~al.}(2007)\citenamefont {Hill},
  \citenamefont {Kalkanci}, \citenamefont {McMurry},\ and\ \citenamefont
  {Koser}}]{Hill2007}%
  \BibitemOpen
  \bibfield  {author} {\bibinfo {author} {\bibfnamefont {J.}~\bibnamefont
  {Hill}}, \bibinfo {author} {\bibfnamefont {O.}~\bibnamefont {Kalkanci}},
  \bibinfo {author} {\bibfnamefont {J.~L.}\ \bibnamefont {McMurry}},\ and\
  \bibinfo {author} {\bibfnamefont {H.}~\bibnamefont {Koser}},\ }\href
  {https://doi.org/10.1103/PhysRevLett.98.068101} {\bibfield  {journal}
  {\bibinfo  {journal} {Phys. Rev. Lett.}\ }\textbf {\bibinfo {volume} {98}},\
  \bibinfo {pages} {068101} (\bibinfo {year} {2007})}\BibitemShut {NoStop}%
\bibitem [{\citenamefont {Kaya}\ and\ \citenamefont {Koser}(2012)}]{Kaya2012}%
  \BibitemOpen
  \bibfield  {author} {\bibinfo {author} {\bibfnamefont {T.}~\bibnamefont
  {Kaya}}\ and\ \bibinfo {author} {\bibfnamefont {H.}~\bibnamefont {Koser}},\
  }\href {https://doi.org/10.1016/j.bpj.2012.03.001} {\bibfield  {journal}
  {\bibinfo  {journal} {Biophys. J.}\ }\textbf {\bibinfo {volume} {102}},\
  \bibinfo {pages} {1514} (\bibinfo {year} {2012})}\BibitemShut {NoStop}%
\bibitem [{\citenamefont {Figueroa-Morales}\ \emph {et~al.}(2015)\citenamefont
  {Figueroa-Morales}, \citenamefont {{Leonardo Mi{\~{n}}o}}, \citenamefont
  {Rivera}, \citenamefont {Caballero}, \citenamefont {Cl{\'{e}}ment},
  \citenamefont {Altshuler},\ and\ \citenamefont
  {Lindner}}]{Figueroa-Morales2015}%
  \BibitemOpen
  \bibfield  {author} {\bibinfo {author} {\bibfnamefont {N.}~\bibnamefont
  {Figueroa-Morales}}, \bibinfo {author} {\bibfnamefont {G.}~\bibnamefont
  {{Leonardo Mi{\~{n}}o}}}, \bibinfo {author} {\bibfnamefont {A.}~\bibnamefont
  {Rivera}}, \bibinfo {author} {\bibfnamefont {R.}~\bibnamefont {Caballero}},
  \bibinfo {author} {\bibfnamefont {E.}~\bibnamefont {Cl{\'{e}}ment}}, \bibinfo
  {author} {\bibfnamefont {E.}~\bibnamefont {Altshuler}},\ and\ \bibinfo
  {author} {\bibfnamefont {A.}~\bibnamefont {Lindner}},\ }\href
  {https://doi.org/10.1039/C5SM00939A} {\bibfield  {journal} {\bibinfo
  {journal} {Soft Matter}\ }\textbf {\bibinfo {volume} {11}},\ \bibinfo {pages}
  {6284} (\bibinfo {year} {2015})}\BibitemShut {NoStop}%
\bibitem [{\citenamefont {Uspal}\ \emph {et~al.}(2015)\citenamefont {Uspal},
  \citenamefont {Popescu}, \citenamefont {Dietrich},\ and\ \citenamefont
  {Tasinkevych}}]{uspal2015rheotaxis}%
  \BibitemOpen
  \bibfield  {author} {\bibinfo {author} {\bibfnamefont {W.~E.}\ \bibnamefont
  {Uspal}}, \bibinfo {author} {\bibfnamefont {M.~N.}\ \bibnamefont {Popescu}},
  \bibinfo {author} {\bibfnamefont {S.}~\bibnamefont {Dietrich}},\ and\
  \bibinfo {author} {\bibfnamefont {M.}~\bibnamefont {Tasinkevych}},\ }\href
  {https://doi.org/https://doi.org/10.1039/C5SM01088H} {\bibfield  {journal}
  {\bibinfo  {journal} {Soft Matter}\ }\textbf {\bibinfo {volume} {11}},\
  \bibinfo {pages} {6613} (\bibinfo {year} {2015})}\BibitemShut {NoStop}%
\bibitem [{\citenamefont {Ishimoto}(2017)}]{ishimoto2017guidance}%
  \BibitemOpen
  \bibfield  {author} {\bibinfo {author} {\bibfnamefont {K.}~\bibnamefont
  {Ishimoto}},\ }\href {https://doi.org/10.1103/PhysRevE.96.043103} {\bibfield
  {journal} {\bibinfo  {journal} {Phys. Rev. E}\ }\textbf {\bibinfo {volume}
  {96}},\ \bibinfo {pages} {043103} (\bibinfo {year} {2017})}\BibitemShut
  {NoStop}%
\bibitem [{\citenamefont {Mathijssen}\ \emph
  {et~al.}(2019{\natexlab{b}})\citenamefont {Mathijssen}, \citenamefont
  {Figueroa-Morales}, \citenamefont {Junot}, \citenamefont {Cl{\'e}ment},
  \citenamefont {Lindner},\ and\ \citenamefont
  {Z{\"o}ttl}}]{mathijssen2019oscillatory}%
  \BibitemOpen
  \bibfield  {author} {\bibinfo {author} {\bibfnamefont {A.~J. T.~M.}\
  \bibnamefont {Mathijssen}}, \bibinfo {author} {\bibfnamefont
  {N.}~\bibnamefont {Figueroa-Morales}}, \bibinfo {author} {\bibfnamefont
  {G.}~\bibnamefont {Junot}}, \bibinfo {author} {\bibfnamefont
  {{\'E}.}~\bibnamefont {Cl{\'e}ment}}, \bibinfo {author} {\bibfnamefont
  {A.}~\bibnamefont {Lindner}},\ and\ \bibinfo {author} {\bibfnamefont
  {A.}~\bibnamefont {Z{\"o}ttl}},\ }\href
  {https://doi.org/10.1038/s41467-019-11360-0} {\bibfield  {journal} {\bibinfo
  {journal} {Nat. Commun.}\ }\textbf {\bibinfo {volume} {10}},\ \bibinfo
  {pages} {3434} (\bibinfo {year} {2019}{\natexlab{b}})}\BibitemShut {NoStop}%
\bibitem [{\citenamefont {Figueroa-Morales}\ \emph
  {et~al.}(2020{\natexlab{a}})\citenamefont {Figueroa-Morales}, \citenamefont
  {Rivera}, \citenamefont {Soto}, \citenamefont {Lindner}, \citenamefont
  {Altshuler},\ and\ \citenamefont {Cl{\'e}ment}}]{figueroa2020coli}%
  \BibitemOpen
  \bibfield  {author} {\bibinfo {author} {\bibfnamefont {N.}~\bibnamefont
  {Figueroa-Morales}}, \bibinfo {author} {\bibfnamefont {A.}~\bibnamefont
  {Rivera}}, \bibinfo {author} {\bibfnamefont {R.}~\bibnamefont {Soto}},
  \bibinfo {author} {\bibfnamefont {A.}~\bibnamefont {Lindner}}, \bibinfo
  {author} {\bibfnamefont {E.}~\bibnamefont {Altshuler}},\ and\ \bibinfo
  {author} {\bibfnamefont {{\'E}.}~\bibnamefont {Cl{\'e}ment}},\ }\href
  {https://doi.org/10.1126/sciadv.aay0155} {\bibfield  {journal} {\bibinfo
  {journal} {Sci. Adv.}\ }\textbf {\bibinfo {volume} {6}},\ \bibinfo {pages}
  {eaay0155} (\bibinfo {year} {2020}{\natexlab{a}})}\BibitemShut {NoStop}%
\bibitem [{\citenamefont {Ren}\ \emph {et~al.}(2017)\citenamefont {Ren},
  \citenamefont {Zhou}, \citenamefont {Mao}, \citenamefont {Xu}, \citenamefont
  {Huang},\ and\ \citenamefont {Mallouk}}]{ren2017rheotaxis}%
  \BibitemOpen
  \bibfield  {author} {\bibinfo {author} {\bibfnamefont {L.}~\bibnamefont
  {Ren}}, \bibinfo {author} {\bibfnamefont {D.}~\bibnamefont {Zhou}}, \bibinfo
  {author} {\bibfnamefont {Z.}~\bibnamefont {Mao}}, \bibinfo {author}
  {\bibfnamefont {P.}~\bibnamefont {Xu}}, \bibinfo {author} {\bibfnamefont
  {T.~J.}\ \bibnamefont {Huang}},\ and\ \bibinfo {author} {\bibfnamefont
  {T.~E.}\ \bibnamefont {Mallouk}},\ }\href
  {https://doi.org/10.1021/acsnano.7b06107} {\bibfield  {journal} {\bibinfo
  {journal} {{ACS} Nano}\ }\textbf {\bibinfo {volume} {11}},\ \bibinfo {pages}
  {10591} (\bibinfo {year} {2017})}\BibitemShut {NoStop}%
\bibitem [{\citenamefont {Palacci}\ \emph {et~al.}(2015)\citenamefont
  {Palacci}, \citenamefont {Sacanna}, \citenamefont {Abramian}, \citenamefont
  {Barral}, \citenamefont {Hanson}, \citenamefont {Grosberg}, \citenamefont
  {Pine},\ and\ \citenamefont {Chaikin}}]{Palacci2015}%
  \BibitemOpen
  \bibfield  {author} {\bibinfo {author} {\bibfnamefont {J.}~\bibnamefont
  {Palacci}}, \bibinfo {author} {\bibfnamefont {S.}~\bibnamefont {Sacanna}},
  \bibinfo {author} {\bibfnamefont {A.}~\bibnamefont {Abramian}}, \bibinfo
  {author} {\bibfnamefont {J.}~\bibnamefont {Barral}}, \bibinfo {author}
  {\bibfnamefont {K.}~\bibnamefont {Hanson}}, \bibinfo {author} {\bibfnamefont
  {A.~Y.}\ \bibnamefont {Grosberg}}, \bibinfo {author} {\bibfnamefont {D.~J.}\
  \bibnamefont {Pine}},\ and\ \bibinfo {author} {\bibfnamefont {P.~M.}\
  \bibnamefont {Chaikin}},\ }\href {https://doi.org/10.1126/sciadv.1400214}
  {\bibfield  {journal} {\bibinfo  {journal} {Sci. Adv.}\ }\textbf {\bibinfo
  {volume} {1}},\ \bibinfo {pages} {e1400214} (\bibinfo {year}
  {2015})}\BibitemShut {NoStop}%
\bibitem [{\citenamefont {Brosseau}\ \emph {et~al.}(2019)\citenamefont
  {Brosseau}, \citenamefont {Usabiaga}, \citenamefont {Lushi}, \citenamefont
  {Wu}, \citenamefont {Ristroph}, \citenamefont {Zhang}, \citenamefont {Ward},\
  and\ \citenamefont {Shelley}}]{brosseau2019relating}%
  \BibitemOpen
  \bibfield  {author} {\bibinfo {author} {\bibfnamefont {Q.}~\bibnamefont
  {Brosseau}}, \bibinfo {author} {\bibfnamefont {F.~B.}\ \bibnamefont
  {Usabiaga}}, \bibinfo {author} {\bibfnamefont {E.}~\bibnamefont {Lushi}},
  \bibinfo {author} {\bibfnamefont {Y.}~\bibnamefont {Wu}}, \bibinfo {author}
  {\bibfnamefont {L.}~\bibnamefont {Ristroph}}, \bibinfo {author}
  {\bibfnamefont {J.}~\bibnamefont {Zhang}}, \bibinfo {author} {\bibfnamefont
  {M.}~\bibnamefont {Ward}},\ and\ \bibinfo {author} {\bibfnamefont {M.~J.}\
  \bibnamefont {Shelley}},\ }\href
  {https://doi.org/10.1103/PhysRevLett.123.178004} {\bibfield  {journal}
  {\bibinfo  {journal} {Phys. Rev. Lett.}\ }\textbf {\bibinfo {volume} {123}},\
  \bibinfo {pages} {178004} (\bibinfo {year} {2019})}\BibitemShut {NoStop}%
\bibitem [{\citenamefont {Baker}\ \emph {et~al.}(2019)\citenamefont {Baker},
  \citenamefont {Kauffman}, \citenamefont {Laskar}, \citenamefont {Shklyaev},
  \citenamefont {Potomkin}, \citenamefont {Dominguez-Rubio}, \citenamefont
  {Shum}, \citenamefont {Cruz-Rivera}, \citenamefont {Aranson}, \citenamefont
  {Balazs},\ and\ \citenamefont {Sen}}]{baker2019fight}%
  \BibitemOpen
  \bibfield  {author} {\bibinfo {author} {\bibfnamefont {R.~D.~C.}\
  \bibnamefont {Baker}}, \bibinfo {author} {\bibfnamefont {J.~E.}\ \bibnamefont
  {Kauffman}}, \bibinfo {author} {\bibfnamefont {A.}~\bibnamefont {Laskar}},
  \bibinfo {author} {\bibfnamefont {O.}~\bibnamefont {Shklyaev}}, \bibinfo
  {author} {\bibfnamefont {M.}~\bibnamefont {Potomkin}}, \bibinfo {author}
  {\bibfnamefont {L.}~\bibnamefont {Dominguez-Rubio}}, \bibinfo {author}
  {\bibfnamefont {H.}~\bibnamefont {Shum}}, \bibinfo {author} {\bibfnamefont
  {Y.}~\bibnamefont {Cruz-Rivera}}, \bibinfo {author} {\bibfnamefont {I.~S.}\
  \bibnamefont {Aranson}}, \bibinfo {author} {\bibfnamefont {A.~C.}\
  \bibnamefont {Balazs}},\ and\ \bibinfo {author} {\bibfnamefont
  {A.}~\bibnamefont {Sen}},\ }\href {https://doi.org/10.1039/C8NR10257K}
  {\bibfield  {journal} {\bibinfo  {journal} {Nanoscale}\ }\textbf {\bibinfo
  {volume} {11}},\ \bibinfo {pages} {10944} (\bibinfo {year}
  {2019})}\BibitemShut {NoStop}%
\bibitem [{\citenamefont {Huang}\ \emph {et~al.}(2019)\citenamefont {Huang},
  \citenamefont {Uslu}, \citenamefont {Katsamba}, \citenamefont {Lauga},
  \citenamefont {Sakar},\ and\ \citenamefont {Nelson}}]{Huang2019}%
  \BibitemOpen
  \bibfield  {author} {\bibinfo {author} {\bibfnamefont {H.-W.}\ \bibnamefont
  {Huang}}, \bibinfo {author} {\bibfnamefont {F.~E.}\ \bibnamefont {Uslu}},
  \bibinfo {author} {\bibfnamefont {P.}~\bibnamefont {Katsamba}}, \bibinfo
  {author} {\bibfnamefont {E.}~\bibnamefont {Lauga}}, \bibinfo {author}
  {\bibfnamefont {M.~S.}\ \bibnamefont {Sakar}},\ and\ \bibinfo {author}
  {\bibfnamefont {B.~J.}\ \bibnamefont {Nelson}},\ }\href
  {https://doi.org/10.1126/sciadv.aau1532} {\bibfield  {journal} {\bibinfo
  {journal} {Sci. Adv.}\ }\textbf {\bibinfo {volume} {5}},\ \bibinfo {pages}
  {eaau1532} (\bibinfo {year} {2019})}\BibitemShut {NoStop}%
\bibitem [{\citenamefont {Najafi}\ and\ \citenamefont
  {Golestanian}(2004)}]{najafi04}%
  \BibitemOpen
  \bibfield  {author} {\bibinfo {author} {\bibfnamefont {A.}~\bibnamefont
  {Najafi}}\ and\ \bibinfo {author} {\bibfnamefont {R.}~\bibnamefont
  {Golestanian}},\ }\href {https://doi.org/10.1103/PhysRevE.69.062901}
  {\bibfield  {journal} {\bibinfo  {journal} {Phys. Rev. E}\ }\textbf {\bibinfo
  {volume} {69}},\ \bibinfo {pages} {062901} (\bibinfo {year}
  {2004})}\BibitemShut {NoStop}%
\bibitem [{\citenamefont {Golestanian}\ and\ \citenamefont
  {Ajdari}(2008)}]{golestanian08}%
  \BibitemOpen
  \bibfield  {author} {\bibinfo {author} {\bibfnamefont {R.}~\bibnamefont
  {Golestanian}}\ and\ \bibinfo {author} {\bibfnamefont {A.}~\bibnamefont
  {Ajdari}},\ }\href {https://doi.org/10.1103/PhysRevE.77.036308} {\bibfield
  {journal} {\bibinfo  {journal} {Phys. Rev. E}\ }\textbf {\bibinfo {volume}
  {77}},\ \bibinfo {pages} {036308} (\bibinfo {year} {2008})}\BibitemShut
  {NoStop}%
\bibitem [{\citenamefont {Dunkel}\ \emph {et~al.}(2010)\citenamefont {Dunkel},
  \citenamefont {Putz}, \citenamefont {Zaid},\ and\ \citenamefont
  {Yeomans}}]{dunkel10}%
  \BibitemOpen
  \bibfield  {author} {\bibinfo {author} {\bibfnamefont {J.}~\bibnamefont
  {Dunkel}}, \bibinfo {author} {\bibfnamefont {V.~B.}\ \bibnamefont {Putz}},
  \bibinfo {author} {\bibfnamefont {I.~M.}\ \bibnamefont {Zaid}},\ and\
  \bibinfo {author} {\bibfnamefont {J.~M.}\ \bibnamefont {Yeomans}},\ }\href
  {https://doi.org/10.1039/C0SM00164C} {\bibfield  {journal} {\bibinfo
  {journal} {Soft Matter}\ }\textbf {\bibinfo {volume} {6}},\ \bibinfo {pages}
  {4268} (\bibinfo {year} {2010})}\BibitemShut {NoStop}%
\bibitem [{\citenamefont {Daddi-Moussa-Ider}\ \emph
  {et~al.}(2018{\natexlab{a}})\citenamefont {Daddi-Moussa-Ider}, \citenamefont
  {Lisicki}, \citenamefont {Mathijssen}, \citenamefont {Hoell}, \citenamefont
  {Goh}, \citenamefont {B{\l}awzdziewicz}, \citenamefont {Menzel},\ and\
  \citenamefont {L{\"o}wen}}]{Daddi2018state}%
  \BibitemOpen
  \bibfield  {author} {\bibinfo {author} {\bibfnamefont {A.}~\bibnamefont
  {Daddi-Moussa-Ider}}, \bibinfo {author} {\bibfnamefont {M.}~\bibnamefont
  {Lisicki}}, \bibinfo {author} {\bibfnamefont {A.~J. T.~M.}\ \bibnamefont
  {Mathijssen}}, \bibinfo {author} {\bibfnamefont {C.}~\bibnamefont {Hoell}},
  \bibinfo {author} {\bibfnamefont {S.}~\bibnamefont {Goh}}, \bibinfo {author}
  {\bibfnamefont {J.}~\bibnamefont {B{\l}awzdziewicz}}, \bibinfo {author}
  {\bibfnamefont {A.~M.}\ \bibnamefont {Menzel}},\ and\ \bibinfo {author}
  {\bibfnamefont {H.}~\bibnamefont {L{\"o}wen}},\ }\href
  {https://doi.org/10.1088/1361-648x/aac470} {\bibfield  {journal} {\bibinfo
  {journal} {J. Phys. Cond. Matt.}\ }\textbf {\bibinfo {volume} {30}},\
  \bibinfo {pages} {254004} (\bibinfo {year} {2018}{\natexlab{a}})}\BibitemShut
  {NoStop}%
\bibitem [{\citenamefont {Nasouri}\ \emph {et~al.}(2019)\citenamefont
  {Nasouri}, \citenamefont {Vilfan},\ and\ \citenamefont
  {Golestanian}}]{nasouri2019efficiency}%
  \BibitemOpen
  \bibfield  {author} {\bibinfo {author} {\bibfnamefont {B.}~\bibnamefont
  {Nasouri}}, \bibinfo {author} {\bibfnamefont {A.}~\bibnamefont {Vilfan}},\
  and\ \bibinfo {author} {\bibfnamefont {R.}~\bibnamefont {Golestanian}},\
  }\href {https://doi.org/10.1103/PhysRevFluids.4.073101} {\bibfield  {journal}
  {\bibinfo  {journal} {Phys. Rev. Fluids}\ }\textbf {\bibinfo {volume} {4}},\
  \bibinfo {pages} {073101} (\bibinfo {year} {2019})}\BibitemShut {NoStop}%
\bibitem [{\citenamefont {Dhont}(1996)}]{dhontBook96}%
  \BibitemOpen
  \bibfield  {author} {\bibinfo {author} {\bibfnamefont {J.~K.~G.}\
  \bibnamefont {Dhont}},\ }\href@noop {} {\emph {\bibinfo {title} {{An
  Introduction to Dynamics of Colloids}}}}\ (\bibinfo  {publisher} {Elsevier,
  Amsterdam},\ \bibinfo {year} {1996})\BibitemShut {NoStop}%
\bibitem [{\citenamefont {Cichocki}\ and\ \citenamefont
  {Jones}(1998)}]{cichocki98}%
  \BibitemOpen
  \bibfield  {author} {\bibinfo {author} {\bibfnamefont {B.}~\bibnamefont
  {Cichocki}}\ and\ \bibinfo {author} {\bibfnamefont {R.~B.}\ \bibnamefont
  {Jones}},\ }\href {https://doi.org/10.1016/S0378-4371(98)00267-2} {\bibfield
  {journal} {\bibinfo  {journal} {Physica A}\ }\textbf {\bibinfo {volume}
  {258}},\ \bibinfo {pages} {273} (\bibinfo {year} {1998})}\BibitemShut
  {NoStop}%
\bibitem [{\citenamefont {Zuk}\ \emph {et~al.}(2014)\citenamefont {Zuk},
  \citenamefont {Wajnryb}, \citenamefont {Mizerski},\ and\ \citenamefont
  {Szymczak}}]{zuk14}%
  \BibitemOpen
  \bibfield  {author} {\bibinfo {author} {\bibfnamefont {P.~J.}\ \bibnamefont
  {Zuk}}, \bibinfo {author} {\bibfnamefont {E.}~\bibnamefont {Wajnryb}},
  \bibinfo {author} {\bibfnamefont {K.~A.}\ \bibnamefont {Mizerski}},\ and\
  \bibinfo {author} {\bibfnamefont {P.}~\bibnamefont {Szymczak}},\ }\href
  {https://doi.org/10.1017/jfm.2013.668} {\bibfield  {journal} {\bibinfo
  {journal} {J. Fluid Mech.}\ }\textbf {\bibinfo {volume} {741}},\ \bibinfo
  {pages} {R5} (\bibinfo {year} {2014})}\BibitemShut {NoStop}%
\bibitem [{\citenamefont {Fossen}(2011)}]{fossen11}%
  \BibitemOpen
  \bibfield  {author} {\bibinfo {author} {\bibfnamefont {T.~I.}\ \bibnamefont
  {Fossen}},\ }\href@noop {} {\emph {\bibinfo {title} {{Handbook of Marine
  Craft Hydrodynamics and Motion Control}}}}\ (\bibinfo  {publisher} {John
  Wiley \& Sons, New Jersey, USA},\ \bibinfo {year} {2011})\BibitemShut
  {NoStop}%
\bibitem [{\citenamefont {Press}(1992)}]{press92}%
  \BibitemOpen
  \bibfield  {author} {\bibinfo {author} {\bibfnamefont {W.~H.}\ \bibnamefont
  {Press}},\ }\href@noop {} {\emph {\bibinfo {title} {{The Art of Scientific
  Computing}}}}\ (\bibinfo  {publisher} {Cambridge university press, Cambridge,
  U.K.},\ \bibinfo {year} {1992})\BibitemShut {NoStop}%
\bibitem [{\citenamefont {Cichocki}\ \emph {et~al.}(2000)\citenamefont
  {Cichocki}, \citenamefont {Jones}, \citenamefont {Kutteh},\ and\
  \citenamefont {Wajnryb}}]{cichocki00}%
  \BibitemOpen
  \bibfield  {author} {\bibinfo {author} {\bibfnamefont {B.}~\bibnamefont
  {Cichocki}}, \bibinfo {author} {\bibfnamefont {R.~B.}\ \bibnamefont {Jones}},
  \bibinfo {author} {\bibfnamefont {R.}~\bibnamefont {Kutteh}},\ and\ \bibinfo
  {author} {\bibfnamefont {E.}~\bibnamefont {Wajnryb}},\ }\href
  {https://doi.org/10.1063/1.480894} {\bibfield  {journal} {\bibinfo  {journal}
  {J. Chem. Phys.}\ }\textbf {\bibinfo {volume} {112}},\ \bibinfo {pages}
  {2548} (\bibinfo {year} {2000})}\BibitemShut {NoStop}%
\bibitem [{\citenamefont {Daddi-Moussa-Ider}\ \emph
  {et~al.}(2018{\natexlab{b}})\citenamefont {Daddi-Moussa-Ider}, \citenamefont
  {Lisicki}, \citenamefont {Hoell},\ and\ \citenamefont {L{\"o}wen}}]{daddi18}%
  \BibitemOpen
  \bibfield  {author} {\bibinfo {author} {\bibfnamefont {A.}~\bibnamefont
  {Daddi-Moussa-Ider}}, \bibinfo {author} {\bibfnamefont {M.}~\bibnamefont
  {Lisicki}}, \bibinfo {author} {\bibfnamefont {C.}~\bibnamefont {Hoell}},\
  and\ \bibinfo {author} {\bibfnamefont {H.}~\bibnamefont {L{\"o}wen}},\ }\href
  {https://doi.org/10.1063/1.5021027} {\bibfield  {journal} {\bibinfo
  {journal} {J. Chem. Phys.}\ }\textbf {\bibinfo {volume} {148}},\ \bibinfo
  {pages} {134904} (\bibinfo {year} {2018}{\natexlab{b}})}\BibitemShut
  {NoStop}%
\bibitem [{\citenamefont {Golestanian}(2008)}]{golestanian08epje}%
  \BibitemOpen
  \bibfield  {author} {\bibinfo {author} {\bibfnamefont {R.}~\bibnamefont
  {Golestanian}},\ }\href {https://doi.org/10.1140/epje/i2007-10276-2}
  {\bibfield  {journal} {\bibinfo  {journal} {Eur. Phys. J. E}\ }\textbf
  {\bibinfo {volume} {25}},\ \bibinfo {pages} {1} (\bibinfo {year}
  {2008})}\BibitemShut {NoStop}%
\bibitem [{\citenamefont {Figueroa-Morales}\ \emph
  {et~al.}(2020{\natexlab{b}})\citenamefont {Figueroa-Morales}, \citenamefont
  {Soto}, \citenamefont {Junot}, \citenamefont {Darnige}, \citenamefont
  {Douarche}, \citenamefont {Martinez}, \citenamefont {Lindner},\ and\
  \citenamefont {Cl\'ement}}]{figueroa2020spatial}%
  \BibitemOpen
  \bibfield  {author} {\bibinfo {author} {\bibfnamefont {N.}~\bibnamefont
  {Figueroa-Morales}}, \bibinfo {author} {\bibfnamefont {R.}~\bibnamefont
  {Soto}}, \bibinfo {author} {\bibfnamefont {G.}~\bibnamefont {Junot}},
  \bibinfo {author} {\bibfnamefont {T.}~\bibnamefont {Darnige}}, \bibinfo
  {author} {\bibfnamefont {C.}~\bibnamefont {Douarche}}, \bibinfo {author}
  {\bibfnamefont {V.~A.}\ \bibnamefont {Martinez}}, \bibinfo {author}
  {\bibfnamefont {A.}~\bibnamefont {Lindner}},\ and\ \bibinfo {author}
  {\bibfnamefont {E.}~\bibnamefont {Cl\'ement}},\ }\href
  {https://doi.org/10.1103/PhysRevX.10.021004} {\bibfield  {journal} {\bibinfo
  {journal} {Phys. Rev. X}\ }\textbf {\bibinfo {volume} {10}},\ \bibinfo
  {pages} {021004} (\bibinfo {year} {2020}{\natexlab{b}})}\BibitemShut
  {NoStop}%
\bibitem [{\citenamefont {Bianchi}\ \emph {et~al.}(2017)\citenamefont
  {Bianchi}, \citenamefont {Saglimbeni},\ and\ \citenamefont
  {Di~Leonardo}}]{bianchi2017holographic}%
  \BibitemOpen
  \bibfield  {author} {\bibinfo {author} {\bibfnamefont {S.}~\bibnamefont
  {Bianchi}}, \bibinfo {author} {\bibfnamefont {F.}~\bibnamefont
  {Saglimbeni}},\ and\ \bibinfo {author} {\bibfnamefont {R.}~\bibnamefont
  {Di~Leonardo}},\ }\href {https://doi.org/10.1103/PhysRevX.7.011010}
  {\bibfield  {journal} {\bibinfo  {journal} {Phys. Rev. X}\ }\textbf {\bibinfo
  {volume} {7}},\ \bibinfo {pages} {011010} (\bibinfo {year}
  {2017})}\BibitemShut {NoStop}%
\bibitem [{\citenamefont {Grosjean}\ \emph {et~al.}(2016)\citenamefont
  {Grosjean}, \citenamefont {Hubert}, \citenamefont {Lagubeau},\ and\
  \citenamefont {Vandewalle}}]{grosjean2016realization}%
  \BibitemOpen
  \bibfield  {author} {\bibinfo {author} {\bibfnamefont {G.}~\bibnamefont
  {Grosjean}}, \bibinfo {author} {\bibfnamefont {M.}~\bibnamefont {Hubert}},
  \bibinfo {author} {\bibfnamefont {G.}~\bibnamefont {Lagubeau}},\ and\
  \bibinfo {author} {\bibfnamefont {N.}~\bibnamefont {Vandewalle}},\ }\href
  {https://doi.org/10.1103/PhysRevE.94.021101} {\bibfield  {journal} {\bibinfo
  {journal} {Phys. Rev. E}\ }\textbf {\bibinfo {volume} {94}},\ \bibinfo
  {pages} {021101(R)} (\bibinfo {year} {2016})}\BibitemShut {NoStop}%
\bibitem [{\citenamefont {Grosjean}\ \emph {et~al.}(2018)\citenamefont
  {Grosjean}, \citenamefont {Hubert},\ and\ \citenamefont
  {Vandewalle}}]{grosjean2018magnetocapillary}%
  \BibitemOpen
  \bibfield  {author} {\bibinfo {author} {\bibfnamefont {G.}~\bibnamefont
  {Grosjean}}, \bibinfo {author} {\bibfnamefont {M.}~\bibnamefont {Hubert}},\
  and\ \bibinfo {author} {\bibfnamefont {N.}~\bibnamefont {Vandewalle}},\
  }\href {https://doi.org/10.1016/j.cis.2017.07.019} {\bibfield  {journal}
  {\bibinfo  {journal} {Adv. Colloid Interf. Sci.}\ }\textbf {\bibinfo {volume}
  {255}},\ \bibinfo {pages} {84} (\bibinfo {year} {2018})}\BibitemShut
  {NoStop}%
\bibitem [{\citenamefont {Dreyfus}\ \emph {et~al.}(2005)\citenamefont
  {Dreyfus}, \citenamefont {Baudry}, \citenamefont {Roper}, \citenamefont
  {Fermigier}, \citenamefont {Stone},\ and\ \citenamefont
  {Bibette}}]{dreyfus2005microscopic}%
  \BibitemOpen
  \bibfield  {author} {\bibinfo {author} {\bibfnamefont {R.}~\bibnamefont
  {Dreyfus}}, \bibinfo {author} {\bibfnamefont {J.}~\bibnamefont {Baudry}},
  \bibinfo {author} {\bibfnamefont {M.~L.}\ \bibnamefont {Roper}}, \bibinfo
  {author} {\bibfnamefont {M.}~\bibnamefont {Fermigier}}, \bibinfo {author}
  {\bibfnamefont {H.~A.}\ \bibnamefont {Stone}},\ and\ \bibinfo {author}
  {\bibfnamefont {J.}~\bibnamefont {Bibette}},\ }\href
  {https://doi.org/10.1038/nature04090} {\bibfield  {journal} {\bibinfo
  {journal} {Nature}\ }\textbf {\bibinfo {volume} {437}},\ \bibinfo {pages}
  {862} (\bibinfo {year} {2005})}\BibitemShut {NoStop}%
\bibitem [{\citenamefont {Miskin}\ \emph {et~al.}(2018)\citenamefont {Miskin},
  \citenamefont {Dorsey}, \citenamefont {Bircan}, \citenamefont {Han},
  \citenamefont {Muller}, \citenamefont {McEuen},\ and\ \citenamefont
  {Cohen}}]{miskin2018graphene}%
  \BibitemOpen
  \bibfield  {author} {\bibinfo {author} {\bibfnamefont {M.~Z.}\ \bibnamefont
  {Miskin}}, \bibinfo {author} {\bibfnamefont {K.~J.}\ \bibnamefont {Dorsey}},
  \bibinfo {author} {\bibfnamefont {B.}~\bibnamefont {Bircan}}, \bibinfo
  {author} {\bibfnamefont {Y.}~\bibnamefont {Han}}, \bibinfo {author}
  {\bibfnamefont {D.~A.}\ \bibnamefont {Muller}}, \bibinfo {author}
  {\bibfnamefont {P.~L.}\ \bibnamefont {McEuen}},\ and\ \bibinfo {author}
  {\bibfnamefont {I.}~\bibnamefont {Cohen}},\ }\href
  {https://doi.org/10.1073/pnas.1712889115} {\bibfield  {journal} {\bibinfo
  {journal} {Proc. Nat. Acad. Sci.}\ }\textbf {\bibinfo {volume} {115}},\
  \bibinfo {pages} {466} (\bibinfo {year} {2018})}\BibitemShut {NoStop}%
\bibitem [{\citenamefont {Cui}\ \emph {et~al.}(2019)\citenamefont {Cui},
  \citenamefont {Huang}, \citenamefont {Luo}, \citenamefont {Testa},
  \citenamefont {Gu}, \citenamefont {Chen}, \citenamefont {Nelson},\ and\
  \citenamefont {Heyderman}}]{cui2019nanomagnetic}%
  \BibitemOpen
  \bibfield  {author} {\bibinfo {author} {\bibfnamefont {J.}~\bibnamefont
  {Cui}}, \bibinfo {author} {\bibfnamefont {T.-Y.}\ \bibnamefont {Huang}},
  \bibinfo {author} {\bibfnamefont {Z.}~\bibnamefont {Luo}}, \bibinfo {author}
  {\bibfnamefont {P.}~\bibnamefont {Testa}}, \bibinfo {author} {\bibfnamefont
  {H.}~\bibnamefont {Gu}}, \bibinfo {author} {\bibfnamefont {X.-Z.}\
  \bibnamefont {Chen}}, \bibinfo {author} {\bibfnamefont {B.~J.}\ \bibnamefont
  {Nelson}},\ and\ \bibinfo {author} {\bibfnamefont {L.~J.}\ \bibnamefont
  {Heyderman}},\ }\href {https://doi.org/10.1038/s41586-019-1713-2} {\bibfield
  {journal} {\bibinfo  {journal} {Nature}\ }\textbf {\bibinfo {volume} {575}},\
  \bibinfo {pages} {164} (\bibinfo {year} {2019})}\BibitemShut {NoStop}%
\bibitem [{\citenamefont {Ge}\ \emph {et~al.}(2013)\citenamefont {Ge},
  \citenamefont {Qi},\ and\ \citenamefont {Dunn}}]{ge2013active}%
  \BibitemOpen
  \bibfield  {author} {\bibinfo {author} {\bibfnamefont {Q.}~\bibnamefont
  {Ge}}, \bibinfo {author} {\bibfnamefont {H.~J.}\ \bibnamefont {Qi}},\ and\
  \bibinfo {author} {\bibfnamefont {M.~L.}\ \bibnamefont {Dunn}},\ }\href
  {https://doi.org/10.1063/1.4819837} {\bibfield  {journal} {\bibinfo
  {journal} {Appl. Phys. Lett.}\ }\textbf {\bibinfo {volume} {103}},\ \bibinfo
  {pages} {131901} (\bibinfo {year} {2013})}\BibitemShut {NoStop}%
\bibitem [{\citenamefont {den Toonder}\ \emph {et~al.}(2008)\citenamefont {den
  Toonder}, \citenamefont {Bos}, \citenamefont {Broer}, \citenamefont
  {Filippini}, \citenamefont {Gillies}, \citenamefont {de~Goede}, \citenamefont
  {Mol}, \citenamefont {Reijme}, \citenamefont {Talen}, \citenamefont
  {Wilderbeek} \emph {et~al.}}]{den2008artificial}%
  \BibitemOpen
  \bibfield  {author} {\bibinfo {author} {\bibfnamefont {J.}~\bibnamefont {den
  Toonder}}, \bibinfo {author} {\bibfnamefont {F.}~\bibnamefont {Bos}},
  \bibinfo {author} {\bibfnamefont {D.}~\bibnamefont {Broer}}, \bibinfo
  {author} {\bibfnamefont {L.}~\bibnamefont {Filippini}}, \bibinfo {author}
  {\bibfnamefont {M.}~\bibnamefont {Gillies}}, \bibinfo {author} {\bibfnamefont
  {J.}~\bibnamefont {de~Goede}}, \bibinfo {author} {\bibfnamefont
  {T.}~\bibnamefont {Mol}}, \bibinfo {author} {\bibfnamefont {M.}~\bibnamefont
  {Reijme}}, \bibinfo {author} {\bibfnamefont {W.}~\bibnamefont {Talen}},
  \bibinfo {author} {\bibfnamefont {H.}~\bibnamefont {Wilderbeek}}, \emph
  {et~al.},\ }\href {https://doi.org/10.1039/B717681C} {\bibfield  {journal}
  {\bibinfo  {journal} {Lab Chip}\ }\textbf {\bibinfo {volume} {8}},\ \bibinfo
  {pages} {533} (\bibinfo {year} {2008})}\BibitemShut {NoStop}%
\bibitem [{\citenamefont {Van~Oosten}\ \emph {et~al.}(2009)\citenamefont
  {Van~Oosten}, \citenamefont {Bastiaansen},\ and\ \citenamefont
  {Broer}}]{van2009printed}%
  \BibitemOpen
  \bibfield  {author} {\bibinfo {author} {\bibfnamefont {C.~L.}\ \bibnamefont
  {Van~Oosten}}, \bibinfo {author} {\bibfnamefont {C.~W.~M.}\ \bibnamefont
  {Bastiaansen}},\ and\ \bibinfo {author} {\bibfnamefont {D.~J.}\ \bibnamefont
  {Broer}},\ }\href {https://doi.org/10.1038/nmat2487} {\bibfield  {journal}
  {\bibinfo  {journal} {Nat. Mater.}\ }\textbf {\bibinfo {volume} {8}},\
  \bibinfo {pages} {677} (\bibinfo {year} {2009})}\BibitemShut {NoStop}%
\bibitem [{\citenamefont {Gu}\ \emph {et~al.}(2020)\citenamefont {Gu},
  \citenamefont {Boehler}, \citenamefont {Cui}, \citenamefont {Secchi},
  \citenamefont {Savorana}, \citenamefont {De~Marco}, \citenamefont
  {Gervasoni}, \citenamefont {Peyron}, \citenamefont {Huang}, \citenamefont
  {Pane} \emph {et~al.}}]{gu2020magnetic}%
  \BibitemOpen
  \bibfield  {author} {\bibinfo {author} {\bibfnamefont {H.}~\bibnamefont
  {Gu}}, \bibinfo {author} {\bibfnamefont {Q.}~\bibnamefont {Boehler}},
  \bibinfo {author} {\bibfnamefont {H.}~\bibnamefont {Cui}}, \bibinfo {author}
  {\bibfnamefont {E.}~\bibnamefont {Secchi}}, \bibinfo {author} {\bibfnamefont
  {G.}~\bibnamefont {Savorana}}, \bibinfo {author} {\bibfnamefont
  {C.}~\bibnamefont {De~Marco}}, \bibinfo {author} {\bibfnamefont
  {S.}~\bibnamefont {Gervasoni}}, \bibinfo {author} {\bibfnamefont
  {Q.}~\bibnamefont {Peyron}}, \bibinfo {author} {\bibfnamefont {T.-Y.}\
  \bibnamefont {Huang}}, \bibinfo {author} {\bibfnamefont {S.}~\bibnamefont
  {Pane}}, \emph {et~al.},\ }\href {https://doi.org/10.1038/s41467-020-16458-4}
  {\bibfield  {journal} {\bibinfo  {journal} {Nat. Comm.}\ }\textbf {\bibinfo
  {volume} {11}},\ \bibinfo {pages} {1} (\bibinfo {year} {2020})}\BibitemShut
  {NoStop}%
\end{thebibliography}
